\documentclass[article, aps, twocolumn, %
    float, nofootinbib%
]{revtex4}

\usepackage{graphicx,times}
\usepackage{natbib}
\usepackage{amssymb,amsmath}

\usepackage[a4paper, %
    left  = 1.91cm, %
    right = 1.91cm, %
    top   = 2.54cm, %
    bottom= 2.54cm %
]{geometry}

\usepackage[colorlinks,%
citecolor=blue, linkcolor=blue]{hyperref}

\allowdisplaybreaks[4]

\begin{document}

\title{\large An implementation of Galactic white dwarf binary data analysis for MLDC-3.1}

\author{Yang Lu}
\author{En-Kun Li}
\thanks{Corresponding author: \href{mailto:lienk@mail.sysu.edu.cn}{lienk@mail.sysu.edu.cn}}
\author{Yi-Ming Hu}
\thanks{Corresponding author: \href{mailto:huyiming@mail.sysu.edu.cn}{huyiming@mail.sysu.edu.cn}}
\author{Jian-dong Zhang}
\author{Jianwei Mei}

\affiliation{%
    \scriptsize MOE Key Laboratory of TianQin Mission, %
    TianQin Research Center for Gravitational Physics %
    \& School of Physics and Astronomy, %
    Frontiers Science Center for TianQin, %
    Gravitational Wave Research Center of CNSA, %
    Sun Yat-sen University (Zhuhai Campus), %
    Zhuhai 519082, China.%
}

\begin{abstract}
    The space-borne gravitational wave detectors will observe a large population of double white dwarf binaries in the Milky Way. 
    However, the search for double white dwarfs in the gravitational wave data will be time-consuming due to the large number of templates involved and antenna response calculation.
    In this paper, we implement an iterative combinatorial algorithm to search for double white dwarfs in MLDC-3.1 data.
    To quickly determine the rough parameters of the target sources, 
    the following algorithms are adopted in a coarse search process:
    (1) using the downsampling method to reduce the number of original data points;
    (2) using the undersampling method to speed up the generation of a single waveform template;
    (3) using the stochastic template bank method to quickly construct the waveform template bank
    while achieving high coverage of the parameter space;
    (4) Combining the FFT acceleration algorithm with the stochastic template bank 
    to reduce the calculation time of a single template.
    A fine search process is applied to further determine the parameters of the signals 
    based on the coarse search, for which we adopt the particle swarm optimization.
    Finally, we detected $\mathcal{O}(10^4)$ double white dwarf signals, 
    validating the feasibility of our method.
\end{abstract}

\maketitle

%
\section{Introduction}           
\label{sect:intro}

The first observation of a gravitational wave (GW) signal, i.e., GW150914, 
was made by Advanced LIGO \cite{LIGOScientific:2016aoc}, 
marking the beginning of a new era in observing the universe with GW.
So far, 90 compact binary mergers have been reported by the LIGO-Virgo-KAGRA collaborations %
\cite{LIGOScientific:2018mvr, LIGOScientific:2020ibl, LIGOScientific:2021usb, LIGOScientific:2021djp}.
With the lower frequency cut-off determined by the Earth's large seismic and gravity-gradient noises~\cite{Hall:2017off,VIRGO:2012dcp,Aso:2013eba}, 
ground-based GW detectors, such as the Advanced LIGO \cite{LIGOScientific:2014pky}, 
Advanced Virgo \cite{VIRGO:2014yos}, and KAGRA \cite{Somiya:2011np}, 
are operated in the $10 - 10^3$ Hz frequency band.
However, there are a large number of GW sources in the millihertz (mHz) band, 
such as
massive black hole binaries~\cite{Katz:2019qlu,Wang:2019ryf},
extreme mass ratio inspirals \cite{Calabrese:2017ypx,Fan:2020zhy}, 
stellar-mass binary black holes~\cite{Sesana:2016ljz,Kyutoku:2016ppx,Liu:2020eko},
double white dwarf (DWD) \cite{Nelemans:2001hp,Yu:2010fq,Breivik:2019lmt,Huang:2020rjf},
and stochastic backgrounds of astrophysical and cosmological origin~\cite{2017LRR....20....2R,Liang:2021bde}.
All these sources can be detected by the proposed space-borne GW detectors, 
e.g. TianQin \cite{TianQin:2015yph} and LISA \cite{LISA:2017pwj}.

In the mHz frequency range, binary star systems are mainly composed of DWD binaries in the Milky Way ( $\mathcal{O}(10^8)$ )~\cite{Nelemans:2001hp}. 
Due to their large number, DWDs are expected to be the most numerous GW sources for space-borne detectors, 
and about ten thousand of DWDs will be detected by LISA and TianQin %
\cite{Lamberts:2018cge,Nelemans:2001hp,Yu:2010fq,Breivik:2019lmt,Huang:2020rjf}.
The detection of GWs from DWDs will significantly improve our understanding of stellar evolution, 
Galactic compact binary systems, 
the distribution of stars in the Milky Way, etc.
More specifically,
(1) DWD mergers are one of the main candidate mechanisms for type Ia Supernovae explosions.
In addition, the near-infrared magnitudes of type Ia Supernovae is considered to be the best ``standard candles''\cite{Barone-Nugent:2012esy}, 
which are of great importance for the study of modern cosmology~\cite{SupernovaCosmologyProject:1998vns,SupernovaSearchTeam:1998fmf}.
(2) DWDs are the end products of the evolution of stellar binaries, 
so the detection of GWs from DWDs can shed light on the formation and evolution of stellar binaries \cite{2014LRR....17....3P,Belczynski:2001uc}.
(3) Due to the mass-transfer, DWDs can form the AM Canum Venaticorum (AM CVn) systems, 
thus, using GW observations in combination with electromagnetic observations can probe the physics of mass transfer processes~\cite{Nelemans:2000es,Marsh:2003rd,2010PASP..122.1133S,Tauris:2018kzq}.
(4) The detached DWDs with short orbital periods (ranging from one hour to a few minutes) are particularly suitable in studying the physics of tides~\cite{Fuller:2012ky,DallOsso:2013uac}.
(5) The overall GW signals from DWDs imprint information about the entire galaxy's stellar population, 
which is helpful in figuring out the structural properties of the Milky Way \cite{Breivik:2019lmt, Benacquista:2005tm, Adams:2012qw, Korol:2018wep, Wilhelm:2020qjc}.

Although the huge amount of DWDs could bring us a wealth of information on GWs, 
its detection also poses great challenges.
First, the superposition of GW signals from DWDs can form a confusion noise in space-borne detectors~\cite{Timpano:2005gm,Huang:2020rjf,Liang:2021bde}. 
That means the received DWDs signals can be confused with each other to the point where individual binaries cannot be resolved~\cite{Crowder:2004ca}. 
Second, about ten thousand DWDs will be resolvable due to either their isolation in frequency space or their relative brightness, 
and how to effectively detect these DWD signals has become a big challenge~\cite{Lamberts:2018cge,Yu:2010fq,Breivik:2019lmt,Huang:2020rjf,Timpano:2005gm}.

To solve this problem and develop LISA data analysis algorithms, 
LISA organized four rounds of mock data challenges from 2006 to 2010, 
called Mock LISA Data Challenge (MLDC)%
\footnote{\url{https://astrogravs.nasa.gov/docs/mldc/}.}
~\cite{MockLISADataChallengeTaskForce:2007iof,MockLISADataChallengeTaskForce:2009wir,Babak:2008aa,Arnaud:2007vr,2007CQGra..24S.551A,Arnaud:2006gm}.
Since 2019, LISA has started a new round of data challenges, named LISA Data Challenges (LDC)%
\footnote{\url{https://lisa-ldc.lal.in2p3.fr.}}.
The amount of DWD signals contained in different rounds of MLDC varies enormously, e.g., 
MLDC-1.1 consists of three single sources data sets and four multiple sources data sets~\cite{Arnaud:2007vr};
MLDC-2.1 contains about 26 million DWD signals~\cite{2007CQGra..24S.551A};
MLDC-3.1 has about 60 million DWD signals~\cite{Babak:2008aa};
MLDC-4 is the ``whole enchilada challenge'', which includes all the sources of MLDC 3.1–3.5 in one data set~\cite{MockLISADataChallengeTaskForce:2009wir}. 

The majority of methods utilized to detect the DWD signals in MLDC data set are based on the matched filtering technique,
which calculates the correlation between the data and the expected waveform, 
and is an optimal strategy in detecting signals in additive, Gaussian, stationary noise~\cite{helstrom1968international}. 
Depending on whether a pre-set bank of templates is required, 
the detection methods can be divided into two categories: stochastic search and grid-based search~\cite{Blaut:2009si}.

For the stochastic search, no template is used for the signal search, 
and a typical implementation involves the Markov Chain Monte Carlo (MCMC) algorithm.
Applying an F-MCMC algorithm, \citet{Cornish:2005qw} resulted in correctly identifying the number and the source parameters of GW signals 
from multiple Galactic binaries within simulated LISA data streams.
Using an extension of the MCMC method named blocked-annealed Metropolis–Hasting (BAM) proposed in Ref.~\cite{Crowder:2007ft}, 
\citet{Littenberg:2011zg} reported about 9\,000 sources in MLDC-4 data.
Applying a Metropolis–Hastings Monte Carlo (MHMC) code, 
the Montana State–JPL (MTJPL) collaboration reported 19\,324 sources\footnote{\url{http://www.tapir.caltech.edu/~mldc/results2/MTJPL-writeup-070618-161814.pdf}} 
in the MLDC-2.1 data set~\cite{MockLISADataChallengeTaskForce:2007iof}.
Adopting the Particle Swarm Optimization (PSO) algorithm with the rejection of spurious sources by cross-validating, 
\citet{Zhang:2021htc} reported 12\,270  sources in LDC 1-4 and 12\,044 sources in MLDC-3.1mod\footnote{The authors adds the GW signals used in MLDC-3.1 to the noise realizations in LDC 1-4.}, respectively.

Different from a stochastic search, which concentrates on the regions with high likelihood~\cite{Blaut:2009si}, 
a grid-based search maps the whole parameter space by constructing grids in parameter space. 
Constructing a template bank in the parameter space needs to balance a number of conflicting constraints:
too dense templates will waste computing resources, and too loose templates will easily miss signals. 
Template-based searches for GWs are often limited by the computational cost associated with searching a large parameter space.
So, it is important to study how to place templates effectively and concisely in the parameter space~\cite{RN262,Prix:2006wm}. 
The problem of constructing a grid is equivalent to the problem of covering $d$-dimensional space with identical hyperellipsoids or overlapping regular lattices~\cite{RN262}.
For example, the hypercubic $\mathbb{Z}_n$ lattice and the $A_n^*$ lattice~\cite{Messenger:2008ta,RN262} are used to construct an efficient template bank, 
which was widely implement in searching for continuous GWs by LIGO \cite{Astone:2010ct, Allen:2005fk, Babak:2006ty, Pisarski:2013dj, Wette:2014tca} and LISA  \cite{Brown:2007se, Blaut:2009zz}.
Using reduced Fisher matrices to build the template grid, \citet{Blaut:2009si} reported 12,805 sources in MLDC-3.1 data.
Another more convenient method to place templates is random or stochastic methods%
\footnote{The largest difference between random and stochastic methods is that whether there are any additional pruning steps.},
which are more effective in high-dimensional parameter space and if the parameter space metric%
\footnote{The metric is defined as a distance measure, which is related to the loss in matched filter SNR for a given template and signal~\cite{Messenger:2008ta}.} 
is non-flat \cite{Messenger:2008ta, Babak:2008rb,Harry:2009ea,Fehrmann:2014cpa, Allen:2022lqr}. 
To make the template coverage more effective, numerous algorithms had been applied to prune the random template bank, e.g.,
remove those which are ``too close" together, adjust their positions, etc.\cite{Harry:2009ea,Fehrmann:2014cpa,Manca:2009xw}.
Throughout the manuscript we call such method as ``stochastic template bank''.

In this paper, we implement an iterative source subtraction method to detect DWDs. 
For quick searches of all the expected candidate signals injected in the ``observation'' data set, 
the following methods will be adopted to speed up the calculation:
(1) the downsampling method is applied to reduce data points~\cite{Blaut:2009si};
(2) the undersampling method is utilized to speed up the generation of waveforms in the template bank~\cite{Zhang:2021htc};
(3) the ``Fast Fourier Transform (FFT)'' algorithm is used to reduce the calculation time of each detection statistic~\cite{Blaut:2009si,Jaranowski:2005hz} ;
(4) stochastic template bank method is implemented to improve the coverage of parameter space.
We call this process as \emph{coarse search}.
Some simplifications are removed in the \emph{fine search} procedure so that the parameters are more precisely recovered.
Based on the candidates provided by the coarse search, we use the PSO algorithm to explore the parameter space.

The paper is organized as follows.
A brief review of relevant background information is presented in Section~\ref{s2}.
The method of downsampling the original data and method of generating waveforms with undersampling method are introduced in Section~\ref{s4}.
Details about the search method and pipeline are discussed in Section~\ref{s5}.
Section~\ref{sec:res_dis} is the search results and discussions.
Section \ref{sec:conclusion} is the conclusion.

\section{Background}
\label{s2}
\subsection{GW Signal of DWD}

In the mHz frequency band, the DWDs are expected to exhibit relatively little frequency evolution.
Thus, the GW strain emitted from a DWD can be safely approximated as (in the source frame) 
\begin{gather}
    h_+ (t) = A_+ \cos\Phi(t) 
    = h_0 \frac{1+\cos^2\iota}{2} \cos\Phi(t), 
    \label{eq:hp}\\ 
    h_\times(t) = A_\times \sin\Phi(t)
    = h_0 \cos\iota \sin\Phi(t), 
    \label{eq:hc}\\
    h_0 = \frac{4 (G \mathcal{M}_c)^{5/3}}{c^4 D_L} (\pi f)^{2/3},
    \label{eq:h0}\\
    \Phi(t) = 2 \pi f t + \pi \dot{f} t^{2} + \phi_{0} ,
    \label{eq:Phi}
\end{gather}
where $+,\times$ represent the two polarization modes of GWs,
$\iota$ is the inclination angle of the quadruple rotation axis with respect to the line of sight (the direction is from the source to the Sun),
$\mathcal{M}_c = (m_1 m_2)^{3/5} / (m_1 + m_2)^{1/5}$ ($m_1$ and $m_2$ are the individual masses of the components of the binary) is the chirp mass of the system,
$D_L$ is the luminosity distance to the source,
$\phi_0$ is the initial phase at the start of the observation,
$f$ and $\dot{f}$ are the frequency of the source and its derivative with respect to time, respectively,
$G$ and $c$ are the gravitational constant and the speed of light, respectively.

Considering the motion of the detectors moving around the Sun, a Doppler modulation of the phase of the waveform should be taken into account, i.e.,
\begin{align}
    \Phi(t) &\rightarrow \Phi(t) + \Phi_D(t), \\
    \Phi_D(t) &= 2\pi (f +\dot{f} t)\, \frac{R}{c} \cos\beta\cos(2\pi f_m t - \lambda),
\end{align}
where $\Phi_D(t)$ is the Doppler modulation, 
$f_m=1/$year is the modulation frequency, 
$\beta$ and $\lambda$ are the latitude and the longitude of the source in ecliptic coordinates, 
$R =1$ AU is the semi-major axis of the guiding center of the satellite constellation.

The measured scalar GW signal at time $t$ by detector channel $\mathcal{I}$ is denoted as $h^\mathcal{I} (t)$.
It is the response of the detector to the GW tensor
\begin{equation}
    h^\mathcal{I}(t) = \sum_{a = +,\times} F^{a, \mathcal{I}} (t; \lambda, \beta, \psi) 
    h_a(t; f, \dot{f}, h_0, \iota, \phi_0),
    \label{eq:response}
\end{equation}
where $F^{a, \mathcal{I}}$ is antenna pattern functions of detector channel $\mathcal{I}$,
$\psi$ is the polarization angle that describes the wave frame with respect to the equatorial coordinate system.

In the present paper, we will use the $\mathcal{F}$-statistic method~\cite{Jaranowski:1998qm} for signal detection, with which
the measured signal can be decomposed into time-dependent and time-independent parts
\begin{equation} 
    h^{\mathcal{I}}(t)  =  \sum_{\mu=1}^{4} 
    \mathcal{A}^{\mu} (h_0, \iota, \phi_0, \psi) 
    h_{\mu}^{\mathcal{I}} (t; f, \dot{f}, \lambda, \beta),
    \label{eq:F_response} 
\end{equation}
where $\mathcal{A}^\mu$ is the signal-amplitudes, which depend only on the four \textit{extrinsic parameters} 
$\{h_0, \iota, \phi_0, \psi\}$, and are independent of the detector $\mathcal{I}$,
\begin{equation}
    \begin{aligned} 
        \mathcal{A}^{1}& = A_{+}\cos\phi_{0}\cos2\psi - A_{\times}\sin\phi_{0}\sin2\psi,\\
        \mathcal{A}^{2} &= A_{+}\cos\phi_{0}\sin2\psi + A_{\times}\sin\phi_{0}\cos2\psi,\\
        \mathcal{A}^{3}& = -A_{+}\sin\phi_{0}\cos2\psi - A_{\times}\cos\phi_{0}\sin2\psi,\\
        \mathcal{A}^{4}& = -A_{+}\sin\phi_{0}\sin2\psi + A_{\times}\cos\phi_{0}\cos2\psi.\\
    \end{aligned}
    \label{eq:four_amp} 
\end{equation}
The four basis waveforms $h^\mathcal{I}_\mu$, which depend only on the four \textit{intrinsic parameters} $ \Theta = \{f, \dot{f}, \lambda, \beta\}$, 
are related to the specifics of the detectors and can be written as%
\footnote{Note that we have suppressed the parameters except for $t$.}
\begin{equation}
    \begin{aligned}
        h^\mathcal{I}_1 (t) = a^\mathcal{I}(t) \cos \phi^\mathcal{I} (t),
        \quad
        h^\mathcal{I}_2 (t) = b^\mathcal{I}(t) \cos\phi^\mathcal{I}(t), 
        \\
        h^\mathcal{I}_3 (t) = a^\mathcal{I}(t) \sin \phi^\mathcal{I}(t),
        \quad
        h^\mathcal{I}_4 (t) = b^\mathcal{I}(t) \sin\phi^\mathcal{I}(t), 
    \end{aligned}
\end{equation}
where $a^\mathcal{I}(t)$ and $b^\mathcal{I}(t)$ are the antenna-pattern functions, 
which depend on the sky position of the DWDs and the features of the detector,
and $\phi^\mathcal{I}(t) = \Phi^\mathcal{I}(t) -\phi_{0}$ is the signal phase at the detector channel $\mathcal{I}$.

\subsection{The Noise Model}

We will focus on the case when the instrumental noise $n(t)$ is assumed to be Gaussian stationary with a zero mean.
Thus, the ensemble average of the Fourier components of the noise $n(f)$ can be written in the following form
\begin{equation}
    \left\langle \tilde{n}(f) \tilde{n}^*(f') \right\rangle = \frac{1}{2} \delta(f-f')S_n(f),
    \label{eq:f_tnoise} 
\end{equation}
where ${}^*$ denotes complex conjugate, and $S_n(f)$ is the single-sided noise power spectral density (PSD)\footnote{This is due to that $n(t)$ is real, $\tilde{n}^*(f) = \tilde{n}(-f)$ and therefore $S_n(-f) = S_n(f)$.}.

Space-borne GW detection suffers from laser phase noise, which can be alleviated through the Time-Delay Interferometry (TDI) technology.
TDI essentially constructs virtually equal-arm interferometers so that the laser phase noise cancels out exactly. 
The three symmetric Michelson channels for each interferometer after TDI are named as channels $X, Y, Z$, 
and the MLDC data are expressed in such form.\cite{1999ApJ...527..814A,Krolak:2004xp}
However, different channels will use the same link, then the instrumental noises in different channels may be correlated with each other.
Considering that all the satellite are identical, we can get one ``optimal" combination by linear combinations of $X$, $Y$, and $Z$ \cite{Prince:2002hp}:
\begin{equation}
    \begin{aligned} 
        A &= \frac{Z - X}{\sqrt{2}}, \\
        E &= \frac{X - 2Y + Z}{\sqrt{6}}, \\
        T &= \frac{X + Y + Z}{\sqrt{3}}.\\ 
    \end{aligned}
    \label{eq:AE_ch}
\end{equation}
In the $A$, $E$, and $T$ channels, the instrumental noise is orthogonal,
and consequently, the noise correlation matrix of these three combinations is diagonal~\cite{Prince:2002hp}.
Thus, in the present paper, if there are no special instructions, the detector channel $\mathcal{I}$ will iterate through $A,~ E$, or $T$.
The details of $h^{A,E,T}_\mu(t)$ can be found in Ref.~\cite{Blaut:2009si}.

For a space-borne GW detector with three satellites forming an approximately equilateral triangle, 
the PSDs of noise for the TDI $A$, $E$ and $T$ channels are~\cite{Estabrook:2000ef,Blaut:2009si} :
\begin{equation}
    \begin{aligned}
        S_n^A(f) = & 
        S_n^E(f) \\
        = & 32 \cos{(\omega L/2)}^2 \sin{(\omega L/2)}^2 
        \\ & 
        \times \big\{ [6+4\cos{(\omega L)} + 2\cos{(2\omega L)}] S^{\rm acc}
            \\ & 
        + [2+\cos{(\omega L)}]S^{\rm opt} \big\},\\
        S_n^T(f) = & 
        128\cos{(\omega L/2)}^2\sin{(\omega L/2)}^4 
        \\ & 
        \times \left[4\sin{(\omega L/2)}^2S^{\rm acc}+S^{\rm opt}\right],
    \end{aligned}
    \label{eq:AE_snr}
\end{equation}
where $\omega = 2\pi f$ is the angular frequency,
$L$ is the arm-length between the satellites,
$S^{\rm acc}$ is the PSD of the  proof-mass noises,
$S^{\rm opt}$ is the PSD of the optical-path noises~\cite{Babak:2008aa} (the parameters used in this article can be found in Appendix \ref{sec:instrument_pars}).

Apart from the instrumental noises, the waveforms of a large number of DWD signals may overlap to create a confusion foreground.
One of the most challenging tasks for the DWD search is to identify individual signals from the foreground.
The unresolved binary systems can form a non-stationary confusion noise that should be added to the overall noise level of space-borne detectors~\cite{Timpano:2005gm}. 
In the process of our analysis, we have included the PSD of the foreground noise, see Appendix \ref{sec:foreground_noise} for more details.

\subsection{Likelihood Function}

Since the true parameters of a GW signal in the time series data $x(t)^{\mathcal{I}}$ are unknown, one can use a waveform $h(t)^{\mathcal{I}}$ which closely mimics the signal, and the residuals $n^{\mathcal{I}}(t) = x^{\mathcal{I}} (t) - h^{\mathcal{I}} (t)$ should be consistent with our model of the instrument noise.
In the case of stationary Gaussian noise, $n^{\mathcal{I}}(t) \sim \mathcal{N} (0, S_n^{\mathcal{I}} )$.
Then, the likelihood function with only the noise in data can be written as \cite{Finn:1992wt,Cutler:2005hc,Prix:2006wm,Allen:2005fk, MockLISADataChallengeTaskForce:2009wir}
\begin{equation}
    P( n^{\mathcal{I}}(t) | S_{n}^{\mathcal{I}})  
    = \kappa \exp\left[ -\frac{1}{2} 
        \left( n^{\mathcal{I}}(t) \mid n^{\mathcal{I}}(t) \right) 
    \right]
    \label{eq:noise_lhfun} 
\end{equation} 
where $\kappa$ is the normalization constant,
the inner product $(\cdot | \cdot)$ is the scalar product which is defined as~\cite{Finn:1992wt, Cutler:1994ys}
\begin{equation}
    (x(t) | y(t))  \equiv  
    4\text{Re} \int_{0}^{+ \infty} 
    \frac{\tilde{x}(f) \tilde{y}^*(f) } {S_n (f)}  df. 
    \label{eq:scalar} 
\end{equation}
where $\tilde{x}(f)$ is the Fourier transform of $x(t)$.

Using Eq.~\eqref{eq:noise_lhfun}, the likelihood of observing data $x^\mathcal{I}(t)$ with a signal $h^\mathcal{I}(t)$ should be expressed as
\begin{equation}
    P \left( x^{\mathcal{I}} | h^\mathcal{I}, S_{n}^{\mathcal{I}} \right)
    = \kappa \exp\left[ -\frac{1}{2}
        \left( x^\mathcal{I} - h^\mathcal{I} |  x^\mathcal{I} -h^\mathcal{I} \right)  
    \right].
    \label{eq:signal_lhfun} 
\end{equation}
Combining Eqs.~\eqref{eq:noise_lhfun} and \eqref{eq:signal_lhfun}, and to put information from all channels into calculation, 
the optimal detection statistic can be given by the likelihood ratio
\begin{equation}
    \begin{aligned}
        \log \Lambda = & 
        \log \prod_{\mathcal{I}} \frac{P( x^{\mathcal{I}}(t)\mid h^\mathcal{I}(t), S_{n}^{\mathcal{I}})}{P( n^{\mathcal{I}}(t)\mid S_{n}^{\mathcal{I}})}  
        \\
        = & 
        \sum_{\mathcal{I}} \bigg ((x^{\mathcal{I}}(t) \mid h^{\mathcal{I}}(t)) - \frac{1}{2}(h^{\mathcal{I}}(t) \mid h^{\mathcal{I}}(t)) \bigg).
        \label{eq:log_lhfun} 
    \end{aligned} 
\end{equation}
where the prodcut/summation implies the assumption that the noise between different channels are independent.

The GW signal produced by the DWD is approximately a monochromatic source.
Thus, over the narrow bandwidth of the signal, we assume the spectral density of the noise can be approximated by a constant, i.e., $S_n(f) \approx S_n(f_0)$, where $f_0$ will be some `central' frequency of the bandwidth or the frequency of the signal \cite{Blaut:2009si,Pisarski:2013dj}.
Employing Parseval's theorem, Eq.~\eqref{eq:scalar} can be approximated by~\cite{Astone:2010ct}
\begin{equation}
    (x(t) | y(t))  \approx  \frac{2}{S_n(f_0)} \int_{0}^{T_0}    x(t) y(t) dt,
    \label{eq:scalar_1} 
\end{equation}
where $T_0$ is the observation time, $x^{\mathcal{I}}(t)$ and $y^{\mathcal{I}}(t)$ are some narrow-band continuous wave signals at frequency $f_0$ or the measured data in the narrow bandwidth around $f_0$.

Using Eq. \eqref{eq:scalar_1} 
the log likelihood ration in Eq. \eqref{eq:log_lhfun} can be written as 
\begin{equation}
    \log \Lambda = \sum_{\mathcal{I}} \frac{2T_0}{S_{n}^{\mathcal{I}}(f_0)}  
    \Big( \left \langle x^{\mathcal{I}}(t)h^{\mathcal{I}}(t) \right \rangle -\frac{1}{2}  \left \langle h^{\mathcal{I}}(t) h^{\mathcal{I}}(t)\right \rangle \Big),
    \label{eq:kog_lhfun_1} 
\end{equation}
due to the stationarity of the noise, here we replace the ensemble average defined in Eq. \eqref{eq:f_tnoise} with time average \cite{Blaut:2009si, Astone:2010ct}
\begin{equation}
    \langle x(t) \rangle = \frac{1}{T_0} \int_0^{T_0} x(t) dt.
    \label{eq:notation_1}
\end{equation}
\subsection{$\mathcal{F}$-statistic}
\label{s3}

The $\mathcal{F}$-statistic method is a matched-filtering detection statistics for continuous GWs, 
which was first introduced by \citet{Jaranowski:1998qm}, 
and subsequently generalized to the multidetector case \cite{Cutler:2005hc,Krolak:2004xp,Cornish:2005qw}. 
In the analysis of continuous GWs, adopting this method can reduce the parameter space to including only the parameters affecting the time evolution of the signal phase \cite{Jaranowski:1998qm, Prix:2006wm}.
This method has been widely utilized in continuous GWs searches, e.g. searches for continuous GWs from spinning neutron stars~\cite{LIGOScientific:2003ijf,LIGOScientific:2006jsu} and DWD signals in MLDC~\cite{MockLISADataChallengeTaskForce:2007iof,MockLISADataChallengeTaskForce:2009wir,Babak:2008aa}.

Assuming there is only one signal contained in the data, then the parameters of a DWD signal can be estimated using the maximum likelihood estimation method.
Combining Eq.~\eqref{eq:F_response}, the log likelihood ratio in Eq.~\eqref{eq:kog_lhfun_1} can be rewritten as
\begin{equation} 
    \log \Lambda = \frac{2T_0}{S_{n}^{\mathcal{I}}(f_0)} 
    \sum_{\mathcal{I}} \bigg(
    \mathcal{A}^{\mu} x^{\mathcal{I}}_{\mu}  - \frac{1}{2}\mathcal{A}^{\mu} \mathcal{M}_{\mu \nu} \mathcal{A}^{\nu} 
    \bigg),
    \label{eq:loglhfun_2} 
\end{equation} 
where the Einstein summation convention is adopted for Greek indices, $\mu, \nu \in[1,4]$ and
\begin{equation}
    \begin{aligned}
        x^{\mathcal{I}}_{\mu} =&
        \sum_{\mathcal{I}} \left \langle x^{\mathcal{I}}(t) h^{\mathcal{I}}_{\mu}(t,\Theta)  \right \rangle, \\
        \mathcal{M}_{\mu \nu} =& 
        \sum_{\mathcal{I}} \left \langle  h^{\mathcal{I}}_{\mu}(t,\Theta)  h^{\mathcal{I}}_{\nu}(t,\Theta)  \right \rangle.
    \end{aligned}
    \label{eq:munu} 
\end{equation}

Analytically maximizing the log likelihood ratio (MLR) over $\mathcal{A}^\mu$, yields the so called $\mathcal{F}$-statistic~\cite{Jaranowski:1998qm}
\begin{equation}
    \mathcal{F} \equiv 
    \mathop{\max}_{\mathcal{A}} \log \Lambda 
    = 
    \frac{T_0}{S_{n}^{\mathcal{I}}(f_0)}  
    \sum_{\mathcal{I}} x^{\mathcal{I}}_{\mu} \mathcal{M}^{\mu \nu} x^{\mathcal{I}}_{\nu},
    \label{eq:F_stat} 
\end{equation}
where $\mathcal{M}^{\mu\nu}$ is the inverse of $\mathcal{M}_{\mu\nu}$, i.e., $\mathcal{M}_{\mu\alpha} \mathcal{M}^{\alpha\nu} = \delta^\nu_\mu$. 
And under the MLR condition, the estimator of $\mathcal{A}^\mu_{\rm MLR}$ is
\begin{equation}
    \mathcal{A}^{\mu}_{\rm MLR} = \sum_\mathcal{I} \mathcal{M}^{\mu\nu} x_\nu^\mathcal{I}.
    \label{eq:A_MLR}
\end{equation}
According to Eq.~\eqref{eq:four_amp}, once the four values of $\mathcal{A}^\mu$ are determined, the values of $\{h_0, \iota, \phi_0, \psi\}$ can be obtained analytically (c.f. appendix \ref{sec:analytical_Amp}).

When the target intrinsic parameters are perfectly matched to the signal, the expectation of the $\mathcal{F}$-statistic (in Eq. \eqref{eq:F_stat}) is~\cite{Cutler:2005hc,Prix:2006wm}.
\begin{equation}
    E[2\mathcal{F}] = 4 + \rho^2,
    \label{eq:expectation_data} 
\end{equation}
where $\rho$ is optimal signal-to-noise ratio (SNR), and $\rho^2 = \sum_\mathcal{I} (h^\mathcal{I}| h^\mathcal{I})$.

\subsection{Acceleration algorithm}
\label{sec:FFT}

According to \citet{Blaut:2009si} for a grid-based search, the FFT algorithm can be used to speed up the calculating of $\mathcal{F}$-statistic in the detecting of GW signals from DWDs in MLDC data set\cite{Blaut:2009si}.
Just as in Ref.~\cite{Blaut:2009si}, combining with Eq.~\eqref{eq:munu}, after some mathematical manipulation, Eq.~\eqref{eq:F_stat} can be rewritten in a more compact form of expression
\begin{equation}
    \begin{aligned}
        \mathcal {F} =& 
        2 \frac{T_0}{S_n^{\mathcal{I}}(f_0)} \sum_{\mathcal{I}} 
        \\ & 
        \frac{\left\{ V\bigl| N^{(u)}\bigr|^2 + U \bigl|N^{(v)}\bigr|^2- 2 \, \mathrm{Re} \left[ W \, N^{(u)} (N^{(v)})^* \right] \right\}}{UV -  \lvert W \rvert ^2 }.
    \end{aligned}
    \label{eq:F_stat_acc}
\end{equation}
where
\begin{equation}
    \begin{aligned}
        U =  2  \left \langle h^{\mathcal{I}}_{1}(t,\Theta)\, h^{\mathcal{I}}_{1}(t,\Theta) \right \rangle, & \quad
        V =  2  \left \langle h^{\mathcal{I}}_{2}(t,\Theta)\, h^{\mathcal{I}}_{2}(t,\Theta) \right \rangle, \\
        Q =  2  \left \langle h^{\mathcal{I}}_{1}(t,\Theta)\, h^{\mathcal{I}}_{2}(t,\Theta) \right \rangle, & \quad
        P =  2  \left \langle h^{\mathcal{I}}_{1}(t,\Theta)\, h^{\mathcal{I}}_{4}(t,\Theta) \right \rangle, \\
    \end{aligned}
    \label{eq:UVQPW}
\end{equation}
are some components of the Matrix $\mathcal{M}_{\mu\nu}$ defined in Eq.~\eqref{eq:munu}, $W = Q + iP$, and
\begin{equation}
    \begin{aligned}
        N^{(u)} =& x_1^{\mathcal{I}} + ix_3^{\mathcal{I}} \\
        = & 
        4\pi f L \sin(2\pi f L) \langle x^{\mathcal{I}}(t) m^{(u)} (t) \exp(i \phi^\mathcal{I}(t)) \rangle, \\
        N^{(v)} =& x_2^{\mathcal{I}} + ix_4^{\mathcal{I}} \\
        = & 
        4\pi f L \sin(2\pi f L) \langle x^{\mathcal{I}}(t) m^{(v)} (t) \exp(i \phi^\mathcal{I}(t)) \rangle,
    \end{aligned}
    \label{eq:NuNv}
\end{equation}
where $m^{(u)}$ and $m^{(v)}$ are complex modulation functions as defined in Ref.~\cite{Blaut:2009si}.
Splitting $\phi(t)$ as $2\pi ft + \phi_{\rm mod}$, where $\phi_\mathrm{mod} =  \pi \dot{f} t^{2} + \Phi_D(t)$, from Eq.~\eqref{eq:F_stat_acc} one can get a general integral part
\begin{align}
    I^{\cal I} = & 
    \int^{T_0}_0 x^{\mathcal{I}}(t) \, m(t;f,\beta,\lambda) \,
    \exp [i \phi_\mathrm{mod} (t;f,\dot{f},\beta,\lambda)] 
    \notag \\ & 
    \times \exp [i 2\pi ft] dt.
    \label{eq:appr_fft} 
\end{align}
This integral will be an FT if both the phase modulation function $\phi_{\rm mod}$ 
and the complex modulation function $m$ are independent of the frequency $f$.

In a narrow frequency band, e.g., 0.1 mHz, 
$\phi_{mod}$ and $m$ can be assumed to be constant and approximately represented with the value at the middle frequency of the band, 
i.e., $f_c$ \cite{Blaut:2009si}.
Consequently, the integral of Eq. \eqref{eq:appr_fft} can be approximated by
\begin{align}
    I^{\cal I} \simeq & 
    \int^{T_0}_0 x^{\mathcal{I}}(t) \, m(t;f_c,\beta,\lambda) \,
    \exp [i \phi_\mathrm{mod} (t;f_c,\dot{f},\beta,\lambda)] 
    \notag \\ & 
    \times \exp [i 2\pi ft] dt.
    \label{eq:appr_fft_l} 
\end{align}
Thus, one can construct the template bank with the nodes of the grid coinciding with the Fourier frequencies, which allows us to compute the $\mathcal{F}$-statistic using the FFT algorithm.

\section{Preparing the Data and GW Template}
\label{s4}

The challenge data set MLDC-3.1~\cite{Prix:2007zh} in which  about $6\times 10^7$  Galactic binaries are buried in is used as the target data set in the present paper.
Among all the signals, $\sim 2.61 \times 10^7$ are detached, which mean that the two components are clearly delineated, separated stars, 
and $\sim 3.42 \times 10^7$ are interacting Galactic binaries, which mean that the two components interact in that there is a mass transfer.
The data set contains a two year long data set ($2^{22}$ samples with 15s sampling) with the first generation of TDI observables $X,~Y,$ and $Z$.
We aim to detect as many as possible of the DWD GW signals from the 40,628 ``bright'' signals\footnote{The ``bright'' signals are the Galactic binaries in the MLDC-3.1 data set, whose signal-to-noise ratio(SNR) is greater than 10 in a single TDI-$X$ channel ~\cite{Babak:2008aa}.}.  

A given DWD system will rotate large number of cycles during the two year period.
This translates to a very small volume in parameter space that any template can cover.
Combined with a large number of target sources, the computational burden can be extraordinary.
In this section, we introduce tricks like downsample and undersample, which can decrease the total search time by roughly two orders of magnitude.

\subsection{Downsampling}

A large number of data points is a big challenge to the calculation.
In order to improve the efficiency of calculation, 
we first apply downsampling to the data \cite{Blaut:2009si}.
Downsampling can be used to reduce the number of data points under the condition of the Nyquist sampling theorem (the sampling rate or the Nyquist rate is equal to twice the upper cutoff frequency of a given signal).
To obtain reasonable data, the following four steps are used to process the data.

\paragraph{Dividing data into small bandwidth}

\begin{figure}[tbp]  
    \centering
    \includegraphics[width=0.8\linewidth]{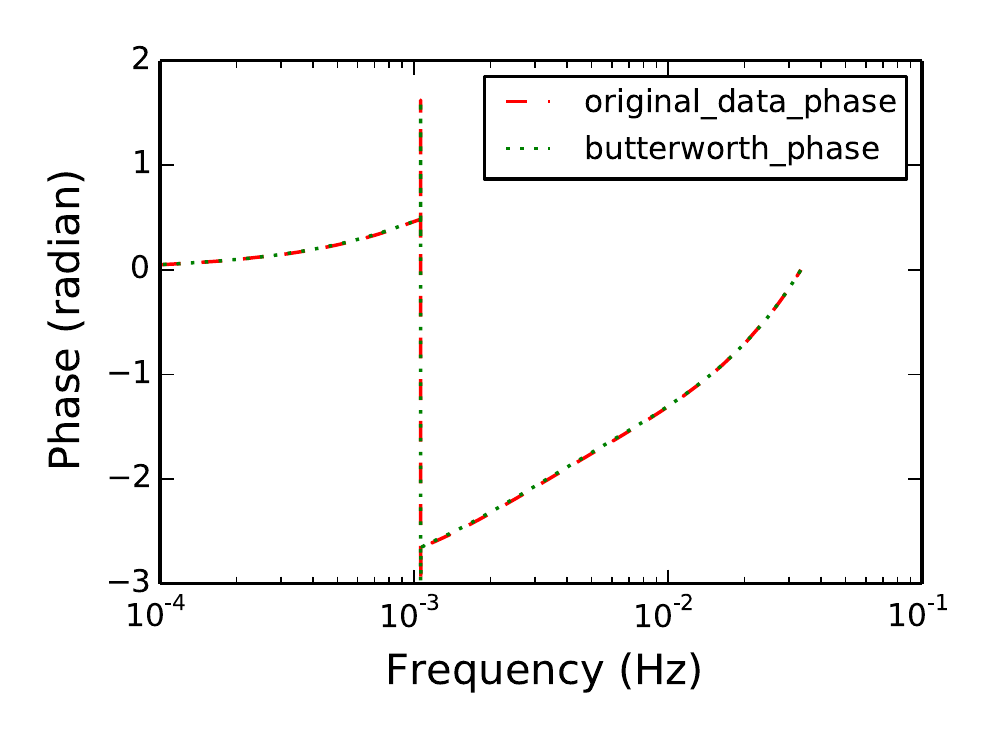}
    \caption{Comparison of signal phase before and after the Butterworth filtering. The frequency of the signal is 1.0627 mHz, 
        and other parameters of the waveform are random.
    Here we adopt the Butterworth filter with bandpass of $[f_1-\epsilon, f_2+\epsilon]$, where $\epsilon=5 \times 10^{-6}$ Hz.}
    \label{Butter_phase}
\end{figure}

During the mission time of the space-borne GW detectors, the evolution of DWD is expected to be small. 
Thus, the GW signals emitted from DWD are nearly monochromatic. 
Therefore, one can split the MLDC-3.1 data into multiple frequency bands, each of which can be analyzed independently.
The segment of each band chosen in this paper is 0.1 mHz. 
To reduce power leakage, the third-order Butterworth bandpass filter is adopted to obtain the narrowband data in the frequency band $[f_1, f_2]$ ($f_2-f_1=0.1$mHz)~\cite{Blaut:2009si}.
The Butterworth filter is applied twice, once forward and once backward. 
As shown in  Fig.~\ref{Butter_phase}, which is a comparison of the phase of the original data and that after the Butterworth filtering, the combined filter introduce zero phase shift.%

\begin{figure}[htbp]
    \centering
    \includegraphics[width=0.8\linewidth]{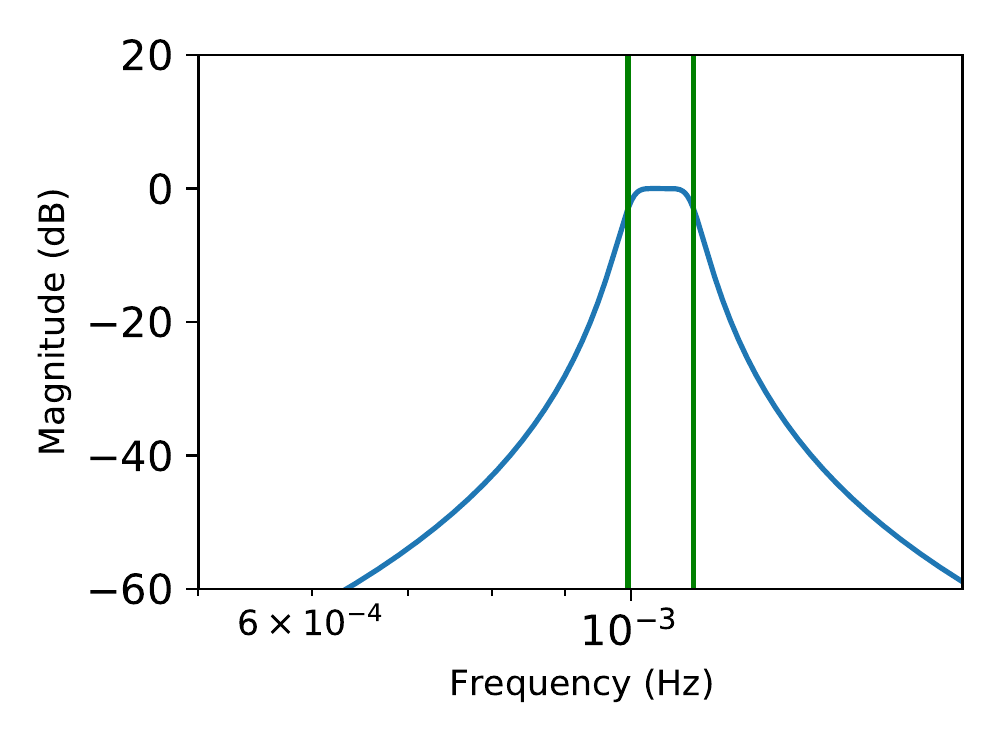}
    \caption{Frequency response of a third-order Butterworth filter for the frequency band of [1.0, 1.1]mHz. }
    \label{Butter_fre_response}
\end{figure}

In Fig.~\ref{Butter_fre_response}, we plot the frequency response of the third-order Butterworth filter with passband $[1.0, 1.1]$ mHz.
One can see that the frequency response in the passband or the stop band is smooth without fluctuations, 
and the stop band attenuation gradually drops to zero. 
The filter can minimize the impact of the filter on the data, as much as possible to preserve the integrity of the filtered data.

\paragraph{Frequency shift}
The bandpassed data can be further downsampled if we hetrodyne the data with a reference monochromatic wave $q(t) = p(t)\cos(2\pi (f_1-\epsilon)t)$.
A signal with original frequency of $f_0$ will be shifted to a lower frequency $(f_0 - (f_1-\epsilon) )$ and a higer frequency $(f_0 + (f_1-\epsilon) )$ components.
Here we have $f_0\in[f_1,f_2]$ and $\epsilon$ is some small number.
More details of the process can be found in Appendix \ref{sec:data_shift}.

\begin{figure}[htbp]
    \centering
    \includegraphics[width=0.8\linewidth]{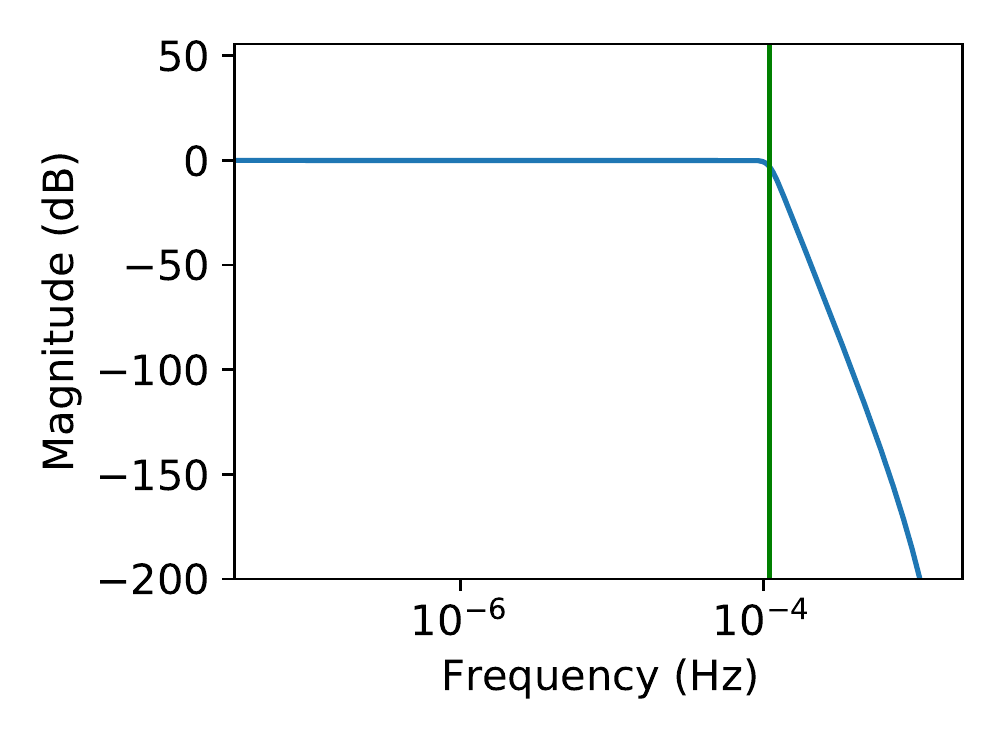}
    \caption{Frequency response of Butterworth filter for lowpass.}
    \label{But_fre_low}
\end{figure}

\paragraph{Lowpass filtering}
We once again use Butterworth filtering for lowpass filtering $[0, f_2 - f_1 + 2\epsilon]$ to filter out the high-frequency data.
The lowpass filter frequency response is shown in Fig. \ref{But_fre_low}.
The Butterworth filter is applied twice: forward and backward in time.

\paragraph{Downsampling}
After the above steps, the center frequency moves from $f_0$ to $(f_0 - (f_1-\epsilon) )$, then we downsample the data under the Nyquist sampling theorem and increase the sampling duration to reduce the amount of data. 

\begin{figure*}[htbp]
    \centering
    \includegraphics[width=0.4\linewidth]{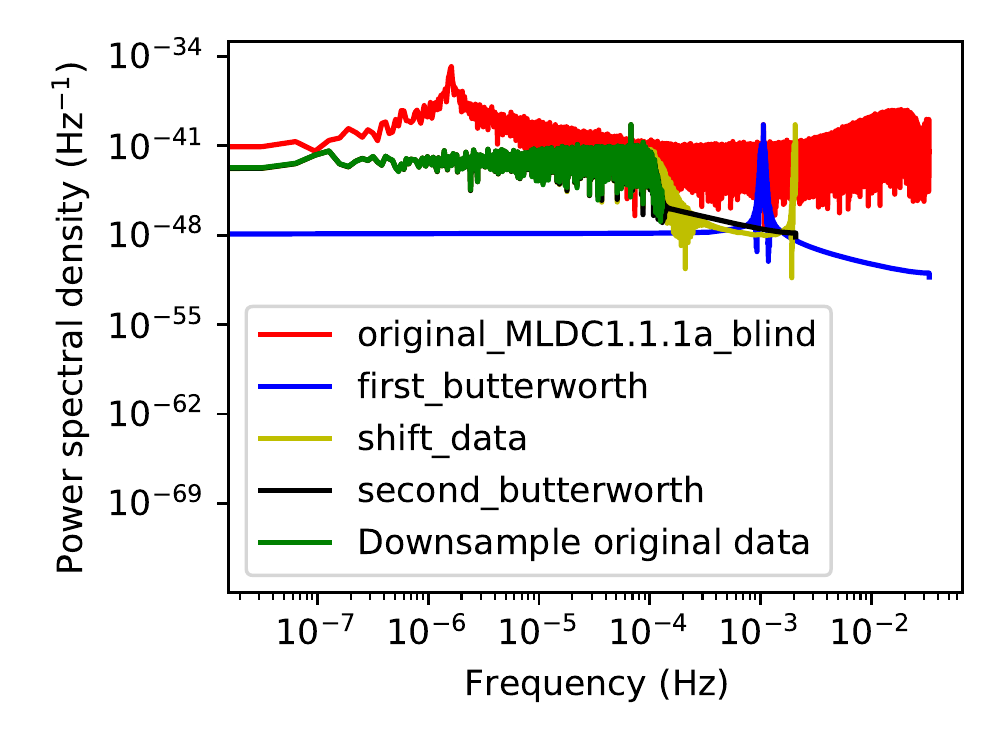}
    \includegraphics[width=0.4\linewidth]{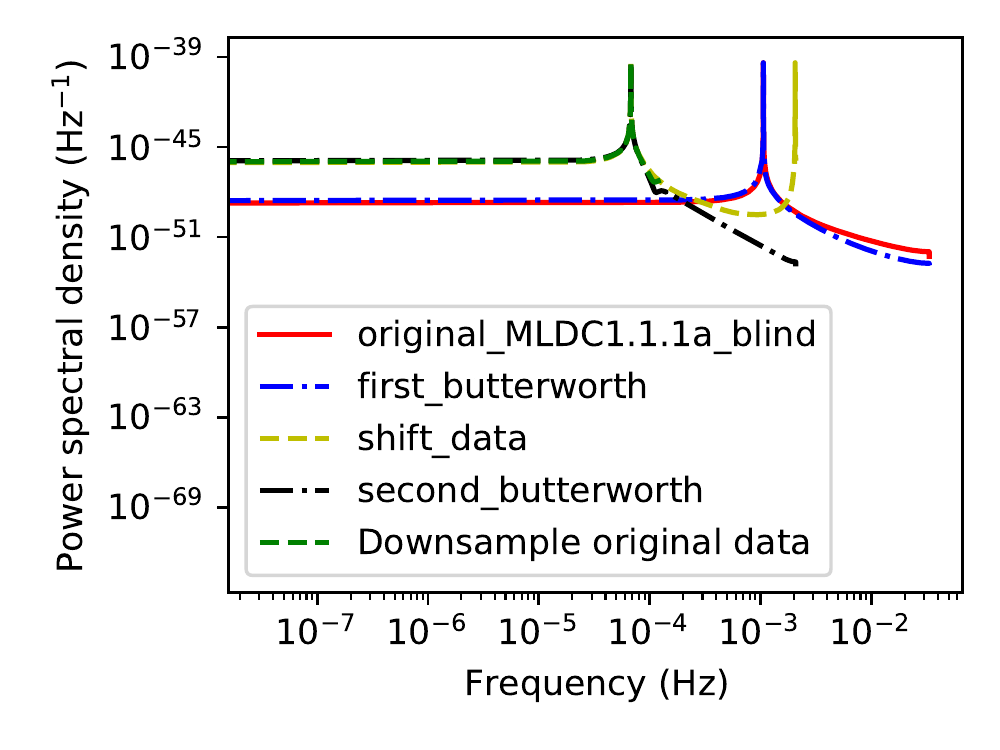}
    \caption{
        The left panel is the relationship between the frequency and PSD of the original data of MLDC1.1.1a-blind (there is only one signal in the data) and the original data after downsampling. 
        The right is the same as the upper picture except that the data is noise-free.
    }
    \label{down_eg}
\end{figure*} 

The data processed through the above four steps is called DS-data.
To test the downsampling method, we have applied it to the MLDC-1.1.1a-blind data.
In Fig.~\ref{down_eg}, the top (down) panel shows  the data of MLDC-1.1.1a-blind(noise-free signal) applying the downsampling method described above.
In the figure, the red line is the original data;
the blue line is the data after bandpass filtering, where the bandpass is $[1\text{mHz} - \epsilon, 1.1\text{mHz} + \epsilon]$, and one can find that the peak of blue line is coincident with the original data;
the yellow line is the data after frequency shift, and one can find that two peaks appear at high and low frequencies, respectively;
the black line is the data after Butterworth lowpass filtering, and the peak at high frequency is filtered out;
the green line is the data after downsampling. 
After the four steps, the number of data points has been reduced by a factor of around $m_d \sim 300$ compare to the original data.

\subsection{Generate Undersampled Waveform Template}

To match the frequencies and the number of the data points in DS-data, the waveform templates should be processed in the same way as the DS-data. 
However, 
applying exactly the same procedures increases computational burden.
The ideal scenario would be process DS-data once, and match with templates generated by downsampling.
This aim can be achieved through undersampled waveform generation.

Undersampling is a technique that samples a bandpass filtered signal at a sampling rate lower than the Nyquist rate, but is still able to reconstruct the signal~\cite{kester2003mixed}.
However, the undersample method may cause aliasing error \cite{Bracewell1986The, kester2003mixed}.
To fully reproduce the original signal, the undersampling rate needs to satisfy the following conditions:
let $f_s' = f_s/m_u$ be the undersampling rate, if $\frac{2f_1}{n} \leq f_s^{\prime} \leq \frac{2f_2}{n-1}$, where integer $n\in (1,\frac{f_2}{f_2-f_1}]$, 
$f_s$ is the original sampling rate, $m_u$ is a positive integer, then the original signal can be fully reproduced in [0, $f_s^{\prime}/2$]. 
Also, a frequency shift like in the previous subsection is used to shift the waveform template to the location of the DS-data.

\begin{figure}[htbp] 
    \centering
    \includegraphics[width=0.8\linewidth]{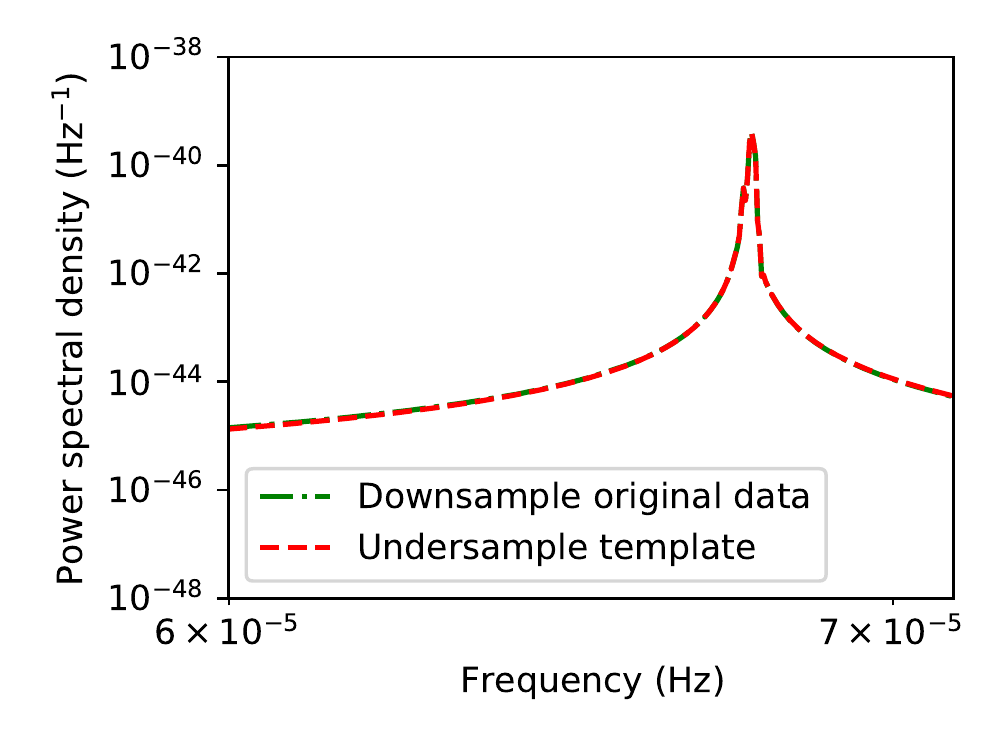}
    \caption{Comparison of the power spectra of the signal downsampling and template undersampling.}
    \label{down_under_wave}
\end{figure}

The data points of the templates obtained after undersampling should be the same as the data points of the data sets that have been downsampled.
Then, the total number of points is reduced by a factor of $m = \min[m_d, m_u]$.
The reduced number of data points can be different in different frequency bands.
In Fig. \ref{down_under_wave}, a comparison is made between the data set after downsampling (i.e., DS-data) and the waveform template generated with undersampling.
To quantify the difference between two signals, one can using the correlation or Fitting Factor (FF)
\begin{equation}
    \mathrm{FF} = \frac{(h_1\mid h_2)}{\sqrt{(h_1 \mid h_1)} \sqrt{(h_2 \mid h_2)} }.
    \label{eq:FF}
\end{equation}
After some calculation, we find that a typical signal would have $\mathrm{FF} \simeq 0.99996$ between the DS-data and the waveform generated with undersampling.

\section{search method}
\label{s5}

Using the methods introduced in the previous section, now we have a DS-data set.
In this section, we describe the construction of template banks. The
search strategy is separated into two parts: 
(1) coarse search, in which we adopt a stochastic template bank method to 
quickly identify the signal candidates; 
(2) fine search, in which we use a PSO algorithm to explore small volumes in parameter space to determine the parameters of the signals.

\subsection{Coarse search} 
\label{sec:coarse_search}

\begin{figure*}[htbp]
    \centering
    \includegraphics[width=1\linewidth]{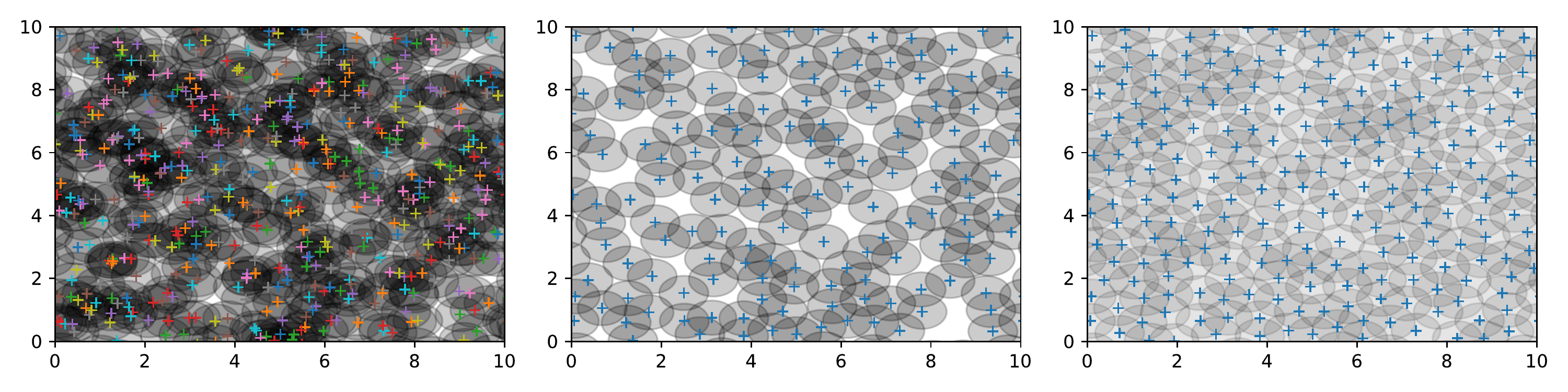}
    \caption{Schematic diagram of stochastic template bank generation in a two-dimensional parameter space, in which the initial coverage is $\eta=0.99$ and the mismatch criteria is $m_*=0.3$.}
    \label{Stochastic_template}
\end{figure*}

To generate the sample data points of the waveform, we need to specify model parameters.
In the present paper, we will choose the template bank search method, thus a target template bank with model parameters needs to be built first.
The template bank construction methods include regular lattice template banks, stochastic bankds and random template banks.
The latter two are expected to have a less coverage than the regular one ~\cite{Messenger:2008ta, Harry:2009ea}.
For the ``random template bank'', the templates are placing randomly with probability distribution determined by the metric.
The ``stochastic template bank'' is similar with the ``random template bank'' but with some additional pruning steps.

The implementation of random template bank method is very simple, 
and can achieve surprisingly high levels of efficiency compared to traditional template banks, 
especially at higher dimensions~\cite{Messenger:2008ta, Allen:2022lqr}.
For a given number of templates, compared with the random template bank method, 
the stochastic template bank will provide better coverage~\cite{Harry:2009ea}.
The stochastic template bank are based on the ``random template bank'', but subtract those templates that are too close,
or perform some other operations (for example, adjust the position according to certain rules)~\cite{Babak:2008rb,Harry:2009ea,Harry:2016ijz,Fehrmann:2014cpa,Allen:2021yuy,VanDenBroeck:2009gd,RN289}.

In this work, we choose the stochastic template bank approach proposed by~\citet{Messenger:2008ta}.
Nearly 100\% parameter space coverage is achieved by subtracting templates that are too close together and then randomly populating some templates, iteratively.
The method can be divided into the following three steps:

\begin{enumerate}
    \item \textit{Randomly generating templates in the parameters space $\mathcal{S}_n$.}
        If the coverage of the templates reachs $\eta$ ($\eta \in [0,1)$) and the mismatch is $m_*$, the number of random templates we need will be $N_R$ \cite{Messenger:2008ta}:
        \begin{equation}
            N_R(\eta, m_*, \mathcal{S}_n) \approx \frac{V_{\mathcal{S}_n}}{V_n} \ln\bigg(\frac{1}{1-\eta}\bigg)m_*^{-n/2}.
            \label{eq:num_rand_temp} 
        \end{equation} 
        where $V_n$  is the volume enclosed by an $n$-dimensional unit sphere, and the proper volume of the parameters space is $V_{\mathcal{S}_n}$.
        
    \item \textit{Removing templates that are too close together.} 
        The distance between two templates at small separation is
        \begin{equation}
            d^2= g_{ij} \Delta\Theta^i \Delta\Theta^{j},
            \label{eq:distance} 
        \end{equation}
        where $g_{ij}$ is the metric of the parameter space.
        Calculate the distance $d_{ij}$ between any two template points, when the two templates are too close, i.e., $d_{ij} < \sqrt{m_*}$, remove one of them.
        Considering that the parameter space is curved, many issues will become complicated, e.g. 
        the distance between widely separated points can no longer be easily computed and the determinant of the metric may be non-constant~\cite{Harry:2009ea}. 
        There are also some efficient stochastic template methods \cite{Fehrmann:2014cpa,Manca:2009xw}.
        Since the efficiency is not dramatically different and here we focus on the implementation of the end-to-end data process pipeline construction.
        Therefore, we stick with the simpler realisation of assume that the metric $g_{ij}$ of two points close to each other is flat 
        ( interested readers are referred to  appendix A of  \cite{Blaut:2009si} for more details). 
    \item Using step 1 to randomly generating about $N_R$ new templates and inject them into the template bank obtained in step 2.
        Then, use step 2 to remove templates that are too close together.
        Repeat the above step until the total number of templates no longer changes \cite{Harry:2009ea}.
        At this time, the coverage of the template bank will be close to 100\%.
\end{enumerate}

To speed up the calculation of distances between every two templates in the template bank, 
in our study, the \texttt{KDTree} algorithm is utilized to speed up the calculation of the distance between templates in steps 2 and 3.
The construction of the template bank takes about 3 days of all frequencies ($f \in [1\times 10^{-4}$,$1.5\times 10^{-2}]$ Hz) for a 3.0 GHz core.
And most of time is used to remove the templates that are too close to each other. 
Note that, the template bank is generated only once. 
An illustration of our entire calculation process of a simple example with two dimensional parameter space is shown in Fig.~\ref{Stochastic_template}.
In this paper, we have set the initial coverage as $\eta = 0.99$ and the mismatch $m_* = 0.3$.
Using the Monte Carlo simulation method as in Ref.~\cite{Messenger:2008ta}, we have computed the spatial coverage in the cases of two-dimensional parameters and obtained a Gaussian distribution with a mean of $\eta = 0.99$.
Following the above three steps, the final coverages only gets better,
which suggest that the stochastic method used in our work is helpful in improving the coverage.

As mentioned in Section~\ref{sec:FFT}, the FFT algorithm can accelerate the calculation while the frequency points of the template bank match the Fourier frequencies.
Thus, the parameter points of $f$ are picked at the Fourier frequencies, and at any frequency point, using the stochastic bank method described above, one can obtain a sub-template bank corresponding to the three parameters, i.e., $\{\lambda, \beta, \dot{f}\}$.
Then, the total number of templates in the template bank for the $i$-th frequency bin will be 
\begin{equation}
    N^i_{\rm total} = N^i_R N^i_{\rm FFT},
\end{equation} 
where $N^i_R$ is the number of the sub-template bank $N_{\rm FFT}$ is the number of points for the FFT.
Note that $N^i_R$ cannot be calculated exactly by Eq.~\eqref{eq:num_rand_temp}, due to that the number of templates removed and added to the template bank is not necessarily the same.

The target frequency band we search in the MLDC-3.1 data is $[1\times10^{-4}, 1.5\times10^{-2}]$ Hz, and we will separate this into 149 frequency bins.
The number of singals above $1.5\times10^{-2}$ Hz is very small and a complete search is not cost-effective.
For the other parameters, we choose  
$\beta \in [-\frac{\pi}{2},\frac{\pi}{2}]$, 
$\lambda\in[0,2\pi]$,
$\dot{f}\in[-3.8\times10^{-17},1.1\times10^{-15}] $ Hz$^2$ for $f<$4 mHz, and 
$\dot{f}\in[-2.3\times10^{-14},7.7\times10^{-14}] $ Hz$^2$ for the rest.

\begin{figure}[htbp]
    \centering
    \includegraphics[width=0.8\linewidth]{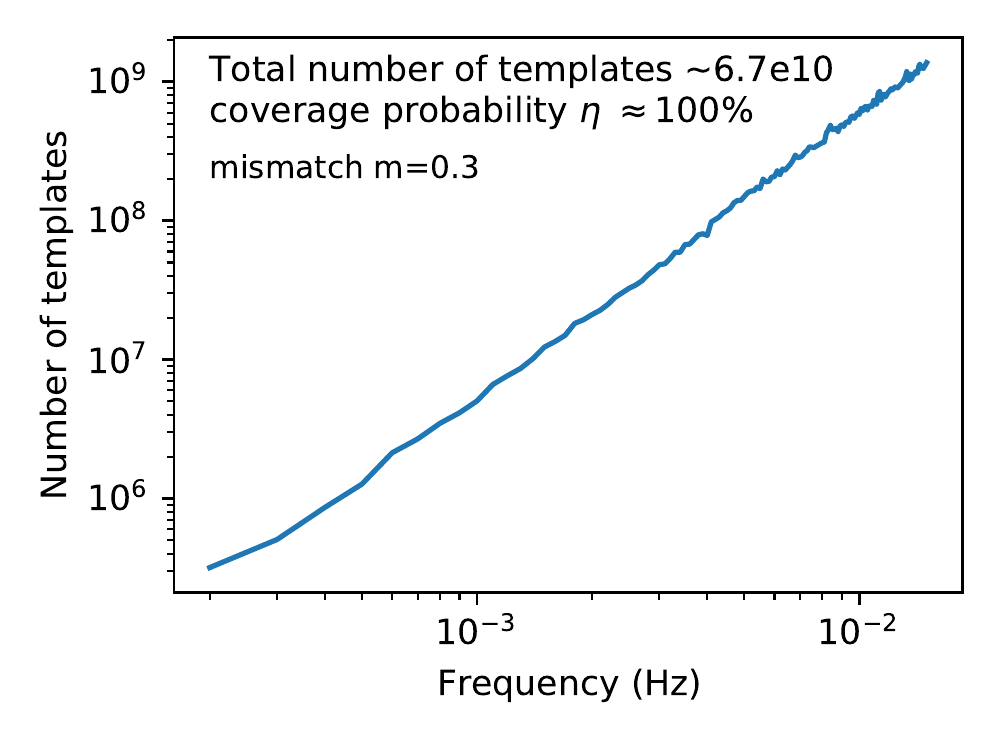}
    \caption{Number of templates required for each frequency band. The total number of template bank for all frequency bands($[1\times10^{-4},1.5\times10^{-2}]$Hz) is about $6.7 \times10^{10}$.}
    \label{num_temp}
\end{figure}

Figure.~\ref{num_temp} plots the number of templates required for different frequency band.
The number of templates increases with frequency, 
and the total number of the templates is about $6 \times 10^{10}$.

In the coarse search, we have downsampled the original MLDC-3.1 data to get the DS-data, 
used the undersampling method to generate the waveform template, and combined the FFT algorithm with stochastic template bank.
The combination of these methods allows us to calculate the $\mathcal{F}$-statistic at a rate of about $10^5 \sim 10^6$ per second. 
Adopting the coarse search, a 3.0 GHz core can search for the $[1\times 10^{-4}, 1.5\times 10^{-2}]$ Hz signal within one day.

\subsection{Fine Search}
\label{sec:PSO}

After the coarse search, the large number of candidate signals obtained was clustered to eliminate redundancy.
In order to refine the determination of match parameters, we adopt the fine search on top of the coarse search. 
In the fine search, one needs to find the maximum of the $\mathcal{F}$-statistic over the parameter space around the clustered templates.

Here we choose the PSO algorithm to do the fine search.
The PSO algorithm is an evolutionary algorithm 
which is based on the concept of swarm intelligence or the simulating of social behavior, 
to solve the search problems in multi-dimensional parameter space.
Previous studies reveal that this algorithm is particularly efficient when trying to find the extrema of multimodal and nontrivial likelihood surfaces~\cite{Bouffanais:2015sya},
and can be more efficient than sophisticated algorithms like MCMC~\cite{Wang:2010jma}.
And PSO algorithms haven been used in astrophysics, such as pulsar timing~\cite{Taylor:2012vx,Wang:2014ava}, ground-based GW astronomy~\cite{Wang:2010jma}, and cosmic microwave background studies~\cite{Prasad:2011rd}.

Here we give a very brief introduction to the PSO algorithm.
In the algorithm, the term ``particle'' is associated with a position $\Theta_i$ (an $n$-dimensional parameter set), and a corresponding velocity $\mathcal{V}_i$ (also defined in the $n$-dimension parameter space). 
$\mathcal{V}_i$ and $\Theta_i$ are $n$-dimension vectors, and $i$ is the index for the $i$-th particle.
A cost function $F_{k}(\Theta_i)$, in this work we adopt the $\mathcal{F}$-statistic, is used to define the fitness of the parameter set to the data.
In the $k$-th iteration, we retain the maximum fit over the $i$-th particle's history $P^k_{i, best}$, as well as the global maximum $g^k_{best}$ among all particles.
The following equations are adopted to update the velocities and positions of the particles
\begin{equation}
    \begin{aligned}
        \mathcal{V}_i^{k+1} &= {\rm w} \mathcal{V}_i^k + c_1 r_1 (P_{i, best}^k - \Theta_i^k) 
        + c_2 r_2 (g_{best}^k - \Theta_i^k) ,\\
        \Theta_i^{k+1} &= \Theta_i^k +\mathcal{V}_i^k.
    \end{aligned}
    \label{eq:sd}
\end{equation}
where $\rm w$ is the inertia weight, $c_1$ and $c_2$ are acceleration constants, and $r_1$, $r_2$ are random numbers drawn uniformly from [0,1]. 
PSO process stops when it converges or reaches the number of iterations and gives a global optimal parameter set $g_{best}$.

For the fine search process in our article, we use the following configuration:
(i) initialize the population parameters:
we adopt the \texttt{scikit-opt} implementation%
\footnote{\url{https://scikit-opt.github.io/scikit-opt/}} 
to perform PSO, where the iterations (\texttt{max\_iter}) is set as 220, and the size of population (\texttt{pop}) is set as 2000.
(ii) Set the parameters' search range: for each cluster 
the ranges of $\lambda, \beta, \dot{f}$ are same as that in coarse search, 
and $f \in [f_c -\Delta f, f_c + \Delta f]$, where $f_c$ and $\Delta f$ are determined by clustering.

For the fine search or the PSO algorithm, it takes about 7 hours for each ``clustered template".
Although there exist faster methods, we adopt the PSO method in the hope that this implementation can serve as a fiducial reference, where we can compare future quicker/smarter algorithms to assess their abilities.

\subsection{Pipeline}
\label{sec:pipeline}

\begin{figure}[htbp]
    \centering
    \includegraphics[width=\linewidth,clip,trim={1.5cm 0.9cm 1.5cm 0.85cm}]{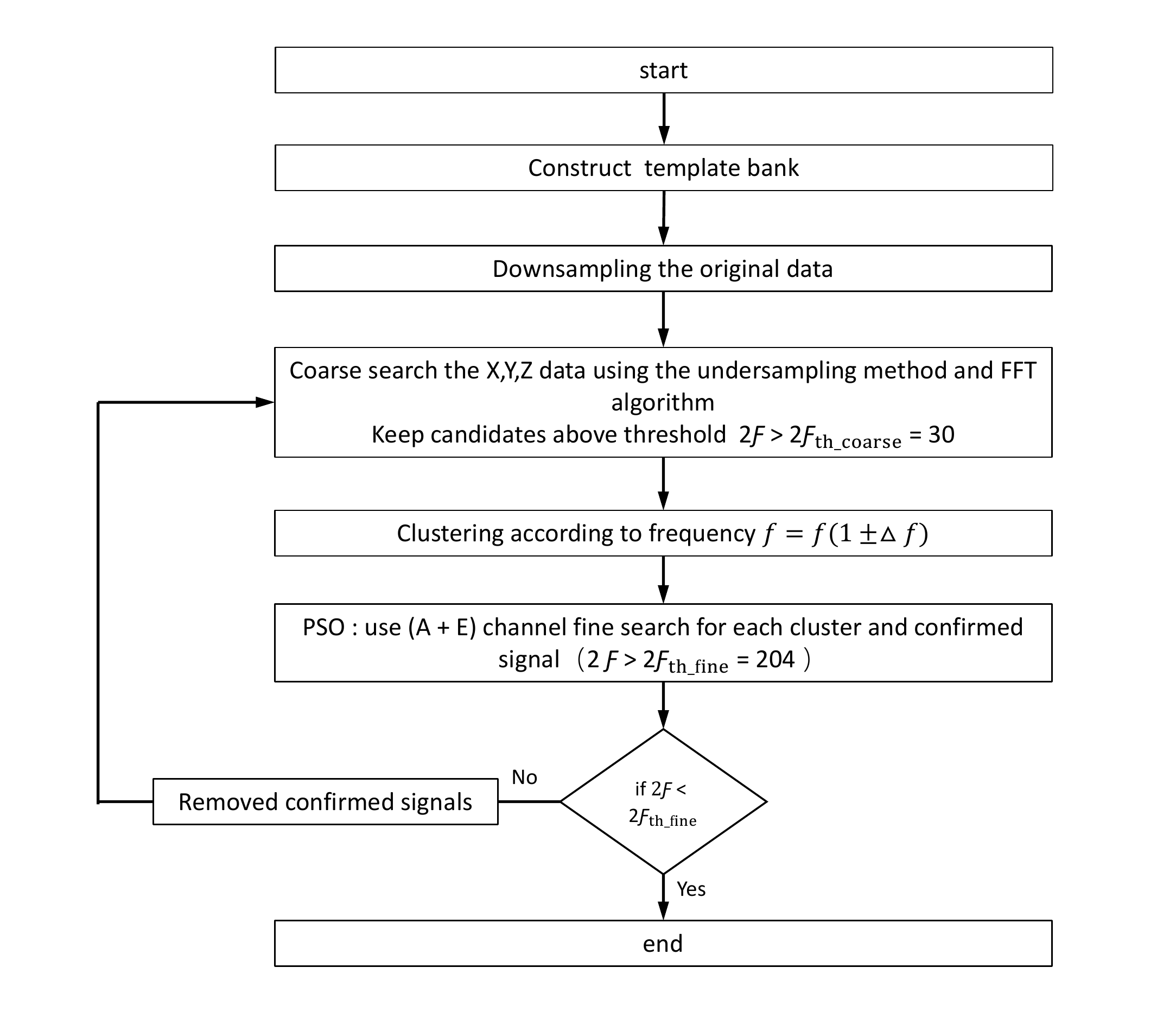}
    \caption{The pipeline we used.}
    \label{pipeline}
\end{figure}

Figure~\ref{pipeline} illustrates the entire search pipeline, including coarse search and fine search. 
The details are described as follows.

First, the coarse search method is adopted to search the TDI channels ($X$, $Y$, $Z$), 
and the candidates with $\mathcal{F}$ above the threshold $\mathcal{F}_{\rm th\_coarse}$ are kept, 
where we have chosen $2 \mathcal{F}_{\rm th\_coarse} = \text{SNR}_{\rm th\_coarse}^2 + 4 = 30$ corresponding to $\text{SNR}_{\rm th\_coarse} \simeq 5.1$.

Second, the candidates that exceed the threshold at different frequency points are clustered together.
It is important to note that due to the motion of the detector around the sun, a Doppler shift will be included within the data.
A real signal of frequency $f_o$ from sources nearly everywhere on the sky will be broadened to the ``double'' Doppler window $\pm 2 \times 10^{-4} f_o $ \cite{Prix:2005gx, Prix:2007zh}. 
The process is as follows:
\begin{itemize}
    \item[(1)] Picking the one with the largest $\mathcal{F}$-statistic among all the candidates and the corresponding frequency is marked as $f_c$. 
    \item[(2)] Finding the candidates whose frequency is between $f_c - \Delta f$ and  $f_c + \Delta f$,  where $\Delta f = 3 \times 10^{-4} f_c$ Hz. 
    \item[(3)] Removing the candidates found in (2) and repeat the previous steps until all candidates have been screened.
\end{itemize}

Third, after clustering, the PSO algorithm is used to do a ``find search'' with the priori provided by each cluster on the DS-data.
The TDI-$A+E$ channel data are used in the fine search.
The parameters are searched in the full range, except for the frequency which is based on the cluster size.
In the fine search, we use the threshold 2$\mathcal{F}_{\rm th\_fine}=204$ equivalently 
${\rm SNR}_{\rm th\_fine} \simeq 14.1$ (the SNR of the A+E channel is $\sqrt{2}$ times higher than that of the single $X,~Y$ or $Z$ channel). 
At the end of fine search, we calculate the SNR with search parameters (consider foreground noise), 
if the SNR of a single $X$ channel is greater than 7, we accept the candidate.
Note that the relationship between $\mathcal{F}$-statistic and SNR is not strict, see Eq.~\eqref{eq:expectation_data}.

Finally, the signal is reconstructed in time domain and removed from the data.
Then, repeat the previous step until there is no more signal whose $\mathcal{F}$-statistic greater than $\mathcal{F}_{\rm th\_fine}$.

\section{Results and Discussions}
\label{sec:res_dis}

Using the pipeline mentioned above, from the MLDC-3.1 blind data set, we have detected 11,519 signals. 
In order to remove the false alarms, we have adopted the following rule:
\begin{itemize}
    \item \it For two detected signals $h_1$ and $h_2$ with frequencies of $f_1$ and $f_2$, if $f_1 -f_2 <1/T_0 = \Delta f$ and the FF between the two signals is $\geq 0.99$, the two signals are considered to be the same, and only the signal with the larger $\mathcal{F}$ is kept.
\end{itemize}
Applying the above criteria, 10,092 signals from 11,521 detected signals had been confirmed.

\subsection{Detection of signals in a single band}
\label{sec:signal_bands}

We first describe the search in a typical frequency band between $[2.5, 2.6]$ mHz.
There are in total of 2,342 bright sources in this frequency band in MLDC-3.1 data set. 
We artificially stop the search when 2$\mathcal{F}_{\rm th\_fine}=204$ in the fine search process.
In this frequency band, we detected 266 signals,
and finally confirmed 209 signals.
Each confirmed signal was paired with all the key signals (40 628 bright Galactic binaries that were injected in MLDC3.1) to calculate FF. 
We only keep maximum corresponding FF for each confirmed signal.
Among all the confirmed signals, there are 156 signals with $FF \geq 0.9$, 28 signals with $ 0.8 \leq FF < 0.9$, and 25 signals with $FF <0.8$. 

\begin{figure}[htbp]
    \centering
    \includegraphics[width=0.8\linewidth]{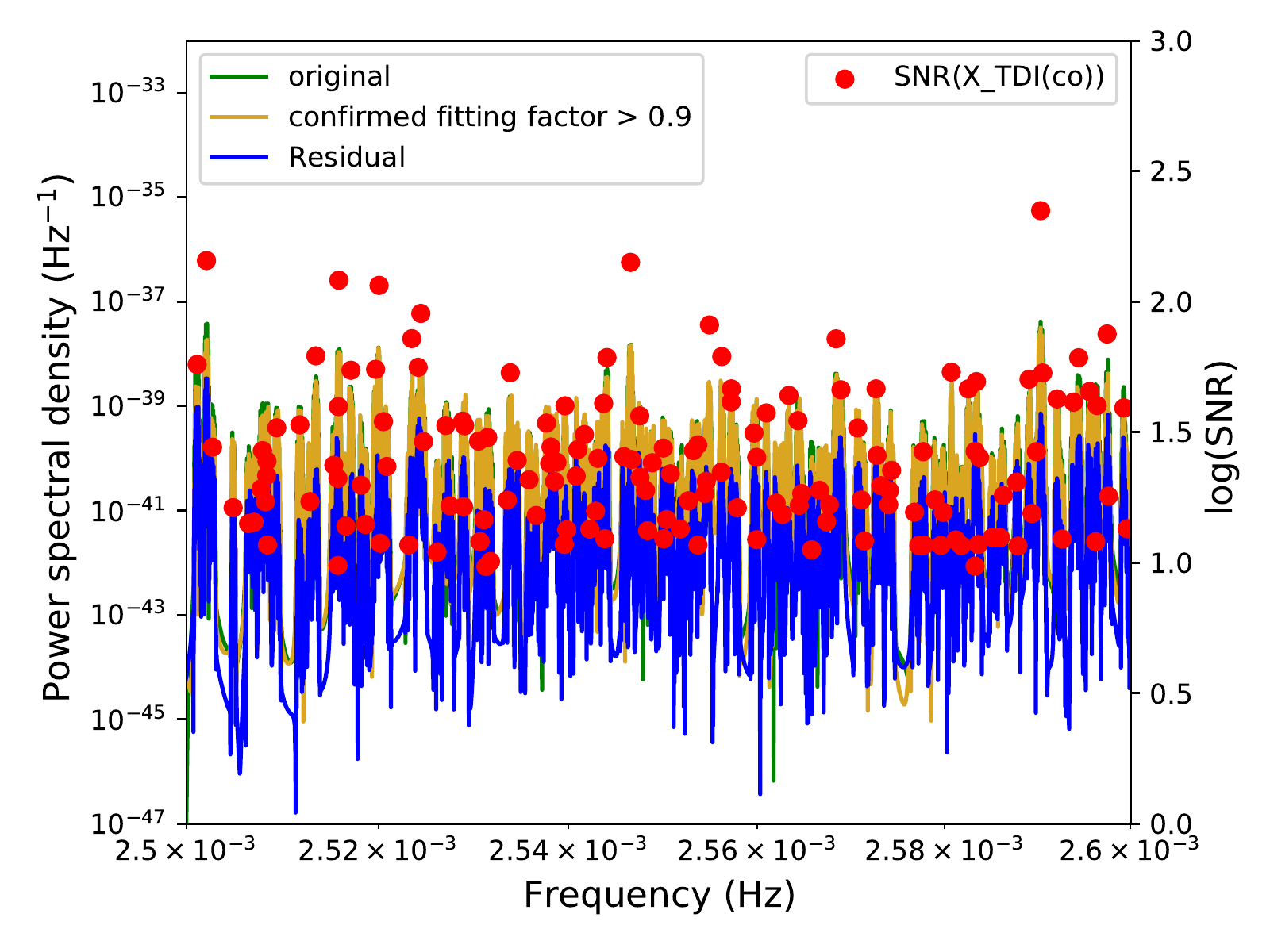}
    \caption{PSDs for the confirmed signals, the MLDC-3.1 real signals, and the reduced data (removing signals) in the range of $[2.5, 2.6]$ mHz band.
    Solid red dots represent the SNR of the confirmed signals.}
    \label{cond_25_26}
\end{figure}

Figure.~\ref{cond_25_26} shows the spectrum of the detected signals and the real signals injected in the MLDC-3.1 data set.
One may note that the residuals of the signals are very large.
Continuing search in this frequency band until no detection statistics having 2$\mathcal{F} > 102$ (SNR $= 7$ corresponding to $X$ channel), we detected an additional 357 signals. 
Among these 357 signals, there are 246 signals with $FF < 0.9$.
This suggests that the residuals are dominated by false signals. 
The ratio of the real signal to the false signal is approximately 1:2. 
This indicates that the signal residuals can have a great impact on the detection of other signals, especially in the low-frequency bands where the signal density is high. 
A better way to cope with the effect of residuals may be using the global-fitting method, e.g. \citet{Littenberg:2020bxy}.

\subsection{Performance of the search method}
\label{sec:performace}

\begin{figure*}[htbp]
    \centering
    \includegraphics[width=0.45\linewidth]{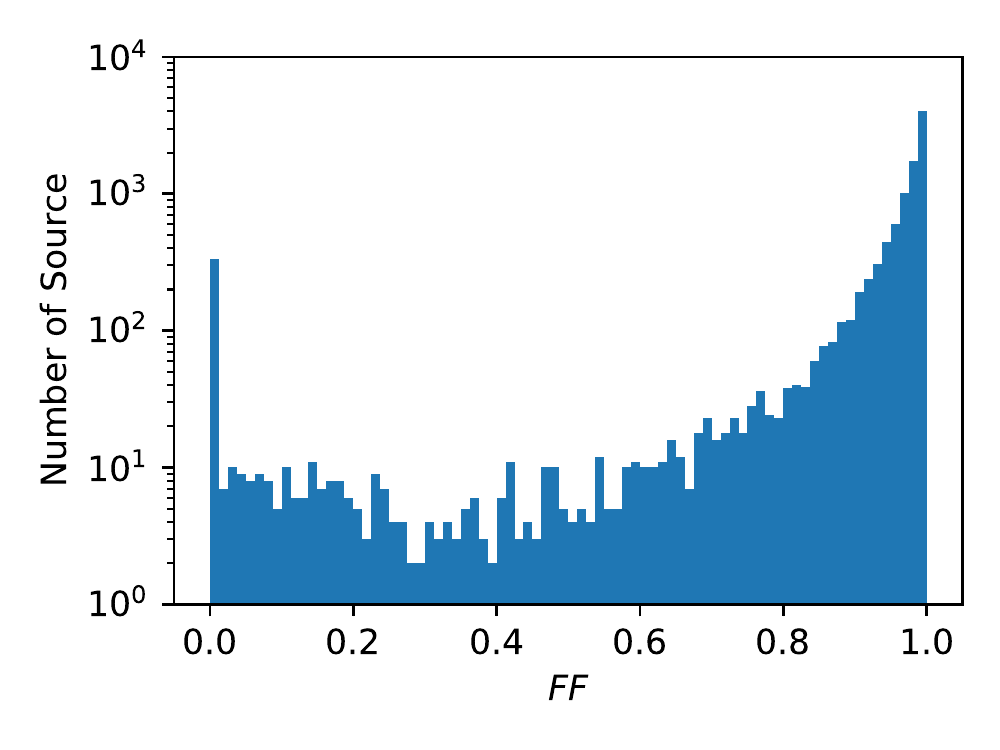}
    \includegraphics[width=0.45\linewidth]{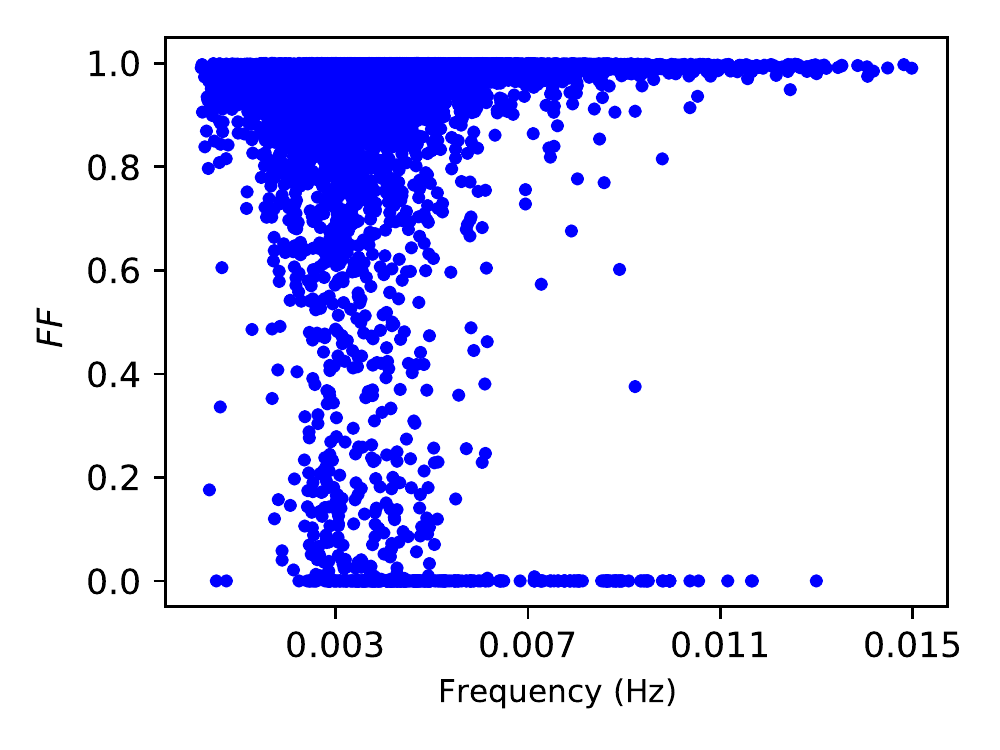}
    \caption{(Left) Histogram of the correlations between our confirmed signals and the injected signals of MLDC-3.1 blind data set.
    (Right) Histogram of the correlations with respect to frequency.}
    \label{fig:FF}
\end{figure*}

Using Eq.~\eqref{eq:FF}, the correlations of the confirmed signals with the injected signals in MLDC-3.1 blind data set are shown in Fig.~\ref{fig:FF}.
In all the confirmed signals, there are 8,600 signals with FF greater than $0.9$ and 1,492 signals with FF less than 0.9.
Among them, there are 573 signals whose FFs are greater than 0.8 while less than 0.9, 107 signals whose FFs are greater than 0.7 while less than 0.8. 
From Fig.~\ref{fig:FF}, one can also find that there is an excess of candidates having correlations $\mathrm{FF} \sim 0$.
Moreover, most of the low FFs originate from the low frequencies, which is consistent with Ref.~\cite{Blaut:2009si}.

\begin{figure*}[htbp]
    \centering
    \includegraphics[width=0.9\linewidth]{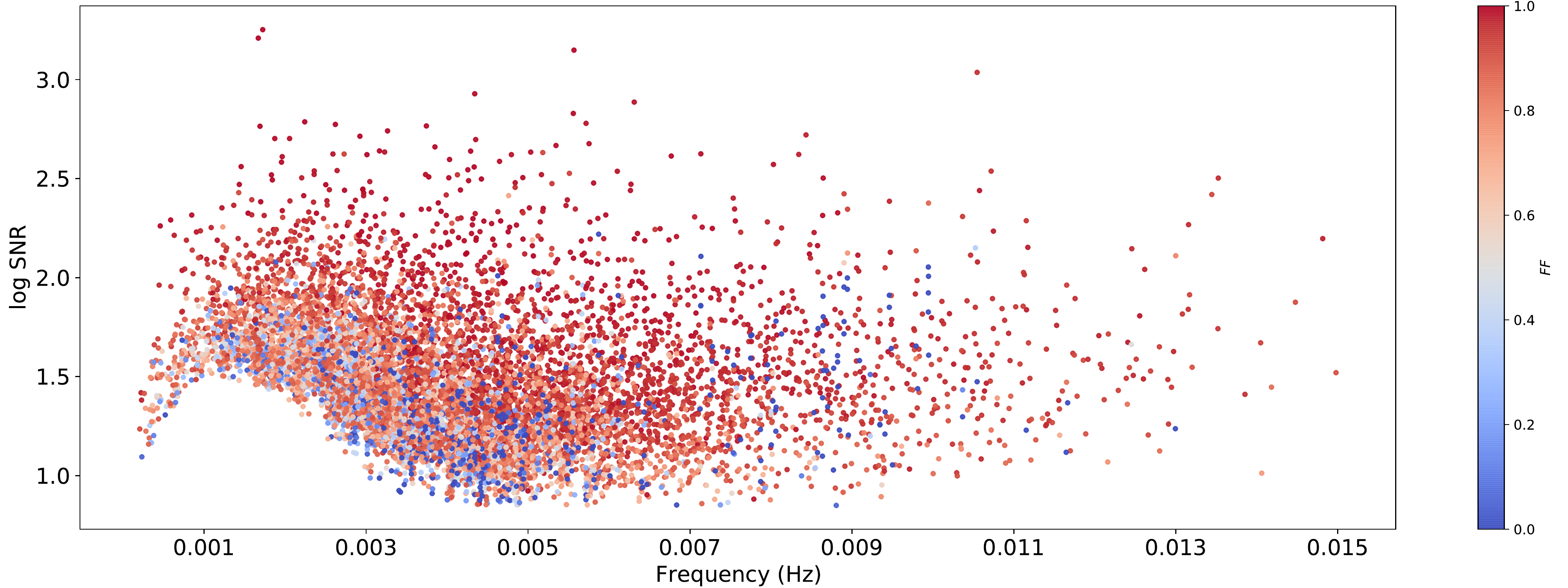}
    \includegraphics[width=0.9\linewidth]{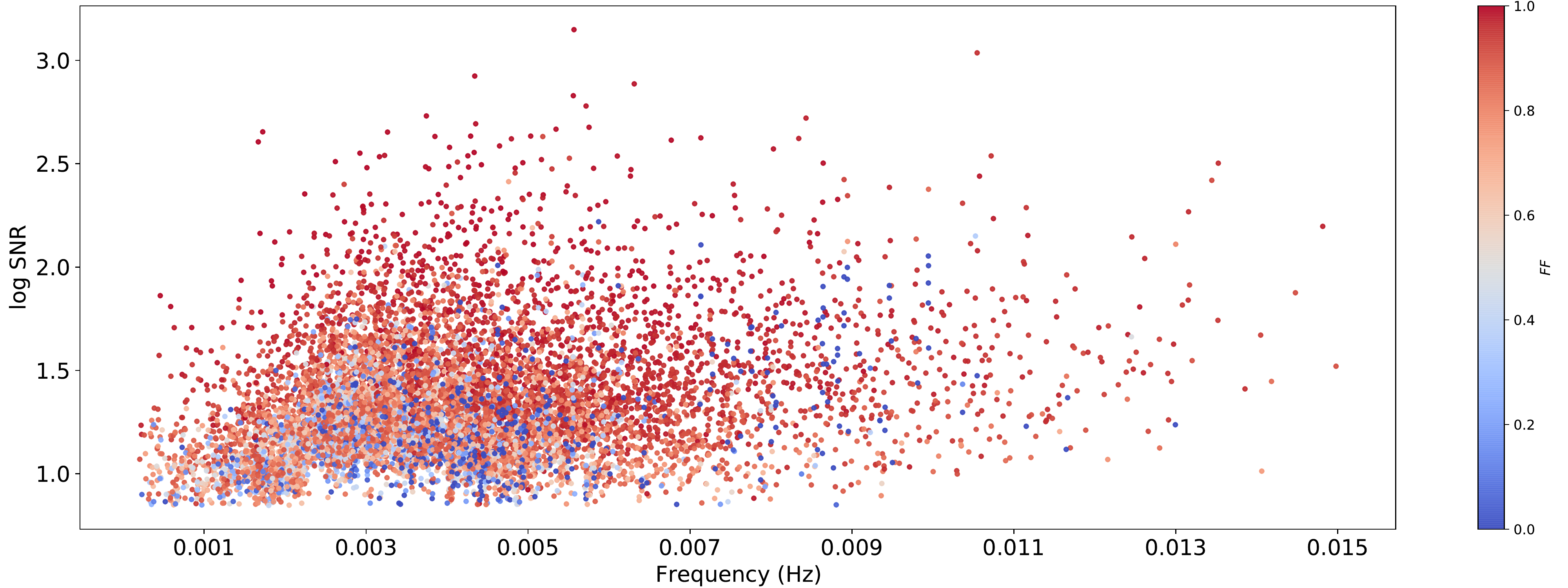}
    \caption{Relationship between the frequency of the confirmed signals and the SNR (top panel: no Galactic confusion noise; bottom panel: with Galactic confusion noise). 
    The color of the dots represents the size of FF.}
    \label{fre_snr_wifor}
\end{figure*}

As described in Ref.~\cite{Blaut:2009si}, the excess of $\mathrm{FF} \sim 0$ signals may be caused by:
(1) the imprecise parameter estimates for some low SNR signals;
(2) many signals with low SNRs interfere with each other causing biases in the parameter estimation.
Figure.~\ref{fre_snr_wifor} shows the relationship between the frequency of all the confirmed signals and the corresponding SNRs.
The color represents the value of the correlated FF, note that these FFs are calculated between our confirmed signals and the released keys of the MLDC-3.1 data set.
In the figure, most blue points are located at low SNRs, i.e., the signals with smaller FFs have mostly smaller SNRs, 
which indicates that the parameter estimation of signals with low SNRs will are poorer.

One can find that in the low frequency range at about $1\sim 3$ mHz, there is a blank area in the upper panel, which suggests that in this low frequency region, even some signals with large SNRs may not be detected.
This is mainly due to the fact that the SNRs shown in the upper panel of Fig.~\ref{fre_snr_wifor} are calculated without the Galactic confusion noise.
When the Galactic confusion noise is taken into consideration, the block will disappear, just as shown in the bottom 
panel of Fig.~\ref{fre_snr_wifor}. 
This indicates that the Galactic confusion noise has a big influence on the detection of the GW signals from DWDs.
And this also indicates that at the low frequency bands, the SNRs of the signals will be reduced due to the presence of Galactic confusion noise.

\begin{figure}[htbp]
    \centering
    \includegraphics[width=0.8\linewidth]{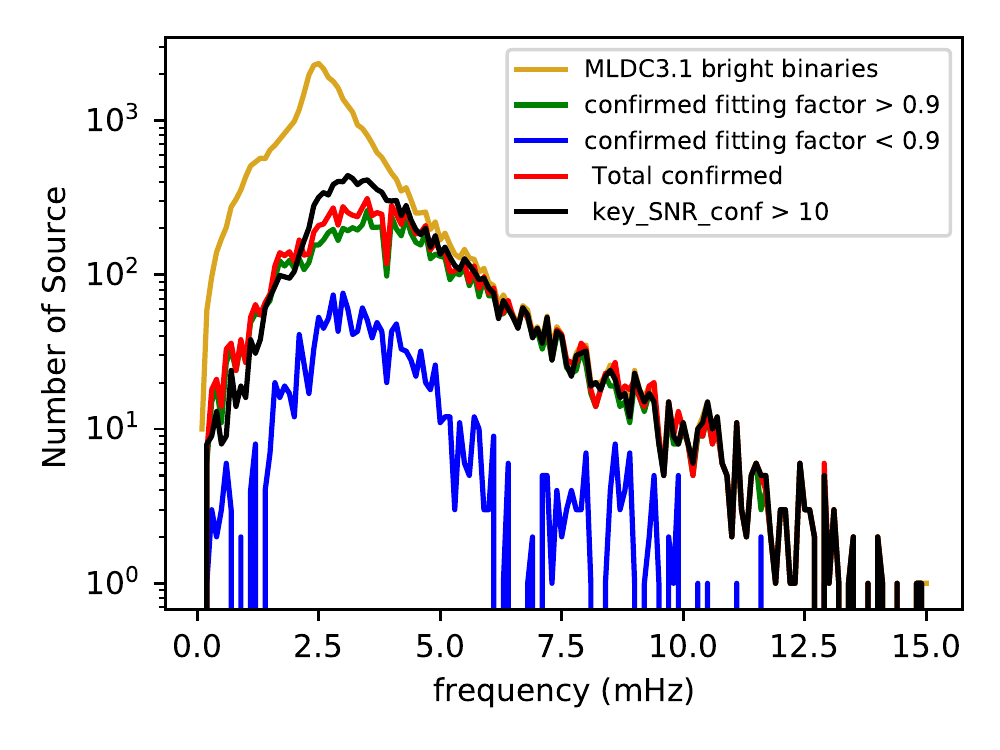}
    \caption{The number of confirmed signals by our search method and the number of signals in MLDC-3.1 bright binaries. 
    }
    \label{num_conf_key}
\end{figure}

To verify this conclusion, the number of bright binaries of MLDC-3.1 and the confirmed signals of this paper in different frequency bands are shown in Fig.~\ref{num_conf_key}.
One can find that the peak value of the bright signals and detection number is not in the same frequency band.
The frequency band of the peak value of the detection number is higher.
There are thousands of signals in each small frequency band around $2.6\times 10^{-3}$ Hz, where only hundreds of signals are confirmed.
This suggests that a large number of signals at low frequencies can interfere with each other and become indistinguishable.
Combining with Fig.~\ref{fre_snr_wifor}, one can see that these unresolvable bright signals form the Galactic confusion noise or foreground noise.

Meanwhile, the black line shown in Fig.~\ref{num_conf_key} is the number of the injected signals whose SNR $>10$ in MLDC-3.1 data set (here, the Galactic foreground noise has been considered).
One can find that the tendency of this line is similar with our confirmed signals especially at high frequency and low frequency, which indicates that our method is trust-worthy.
And the largest deviation appears around $f \sim 3$ mHz, suggest that the removing of large number of signals will leave large residuals and have great impacts on the detection of the remain signals.

Due to the presence of the foreground, the small SNRs prevent 
us from deepening the search sensitivity, the rising portion of false alarms made the further search meaningless.
At high frequencies, the aliasing of signals are significantly reduced, and the detection results are gradually improve.

\subsection{Performance of parameter estimation}
\label{sec:errors}

\begin{figure*}[htbp]
    \centering
    \includegraphics[width=1\linewidth]{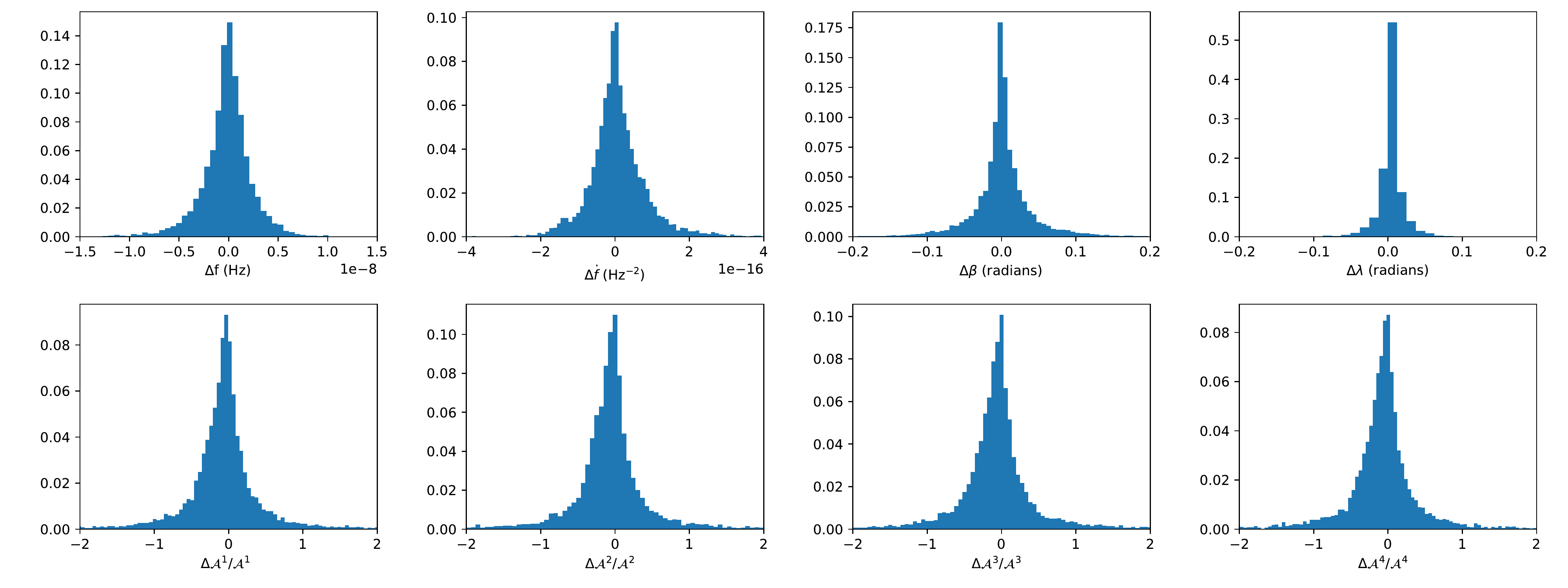}
    \caption{The error distribution between the source parameters we confirmed and the true source parameters in MLDC-3.1 data set. }
    \label{error_parma}
\end{figure*}

Any systematic biases in the parameter estimation can be revealed by displaying the distribution of errors for all the confirmed signals\cite{Littenberg:2011zg}.
We define the parameter errors and the fractional parameter error as \cite{Babak:2008aa}
\begin{align} 
    \Delta \lambda &= \lambda_{\rm rec} - \lambda_{\rm key},
    \label{eq:error}\\ 
    \Delta \lambda / \lambda &=  (\lambda_{\rm rec} - \lambda_{\rm key}) /  \lambda_{\rm key},
    \label{eq:fraction} 
\end{align}
where $\lambda_{\rm rec}$ is the parameter of the confirmed signals, 
$\lambda_{\rm key}$ is the parameter of the injected signals of the MLDC-3.1 blind data. 
Histograms in Fig.~\ref{error_parma} show the distribution of errors for all the signals we confirmed.
Most of the frequency errors are within a small fraction of a Fourier bin ($\Delta f = 1/T_0$), while the errors in frequency derivative are within $\Delta f^2$, and the errors in sky position are within $\pm 0.05$ radians.
In addition, the distributions of all the parameters show a strong peak at zero bias, which indicates that our results are reasonable.

\subsection{Residuals}
\label{sec:residuals}

\begin{figure}[htbp]
    \centering
    \includegraphics[width=1\linewidth]{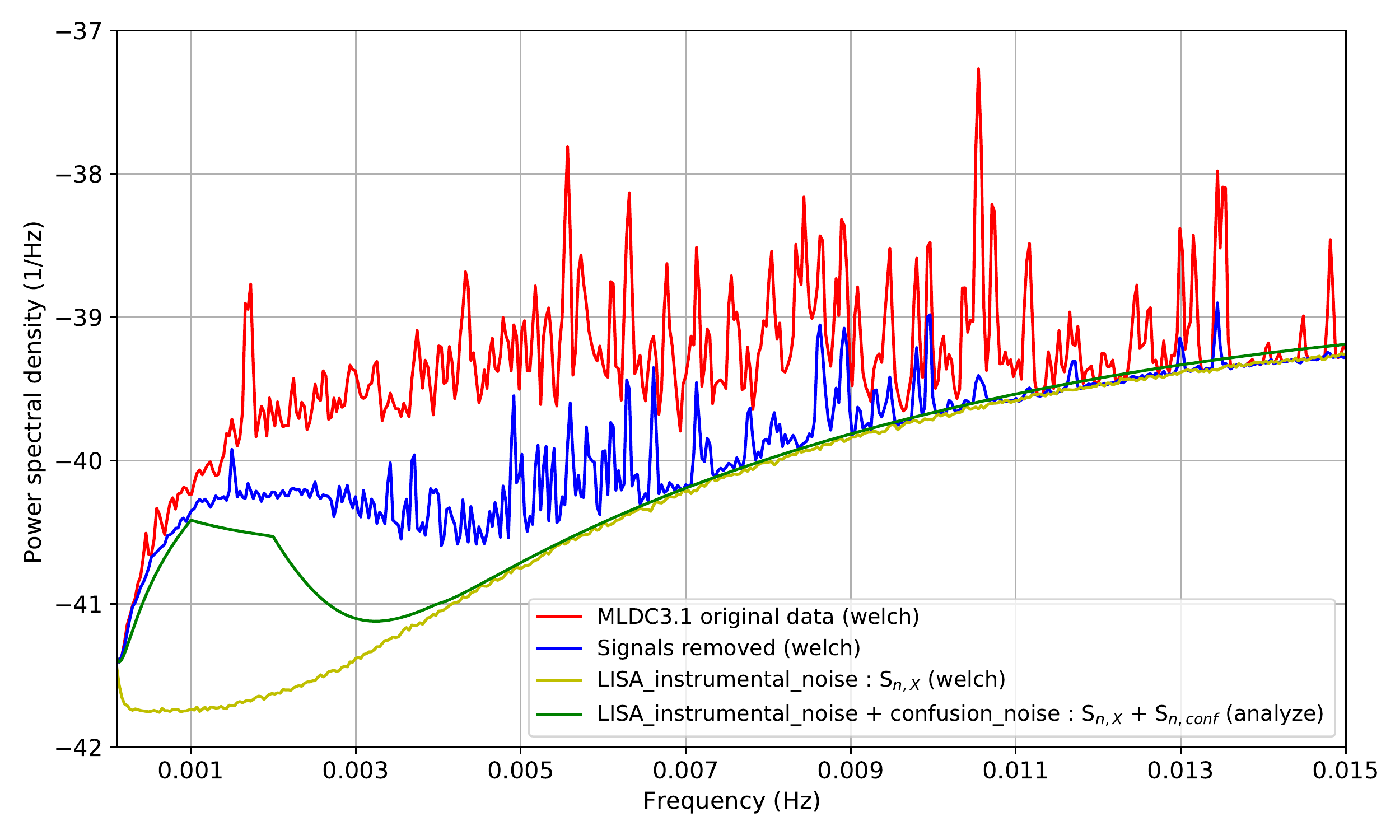}
    \caption{The red line is the smoothed PSD for the original MLDC-3.1 data, the blue line is that of reduced data after removing all confirmed signals, the yellow line is that of the LISA instrumental noise, and the green one is the analytical total noise PSD. 
    }
    \label{psd_red_instr}
\end{figure}

The ability of the signal search can be reflected in another way by comparing the remaining residuals with the noise.
In Fig.~\ref{psd_red_instr} we have compared the smoothed spectrum of the MLDC-3.1 data set with that of data with confirmed signals removed. 
The two smoothed PSDs are compared with the PSD of the LISA instrumental noise (yellow line) and the analytical PSD of instrumental noise with the Galactic confusion noise (green line).
Similar to the results shown in Ref.~\cite{Blaut:2009si}, one can also conclude that above frequency of about 6 mHz, all the DWD systems are resolved well.

Another noteworthy issue is that we use an analytical PSD of instrument noise plus the analytical PSD of Galactic confusion noise (i.e., Eq.~\eqref{eq:confusion}) as the total PSD of noise when searching for a signal.
As shown in Fig.~\ref{psd_red_instr}, in the low frequency bands, the convex part of the analytical total noise PSD matched well with the remaining residuals.
This is good evidence that our search method performs well for searching signals in low frequencies.

\begin{figure}[htbp]
    \centering
    \includegraphics[width=0.8\linewidth,]{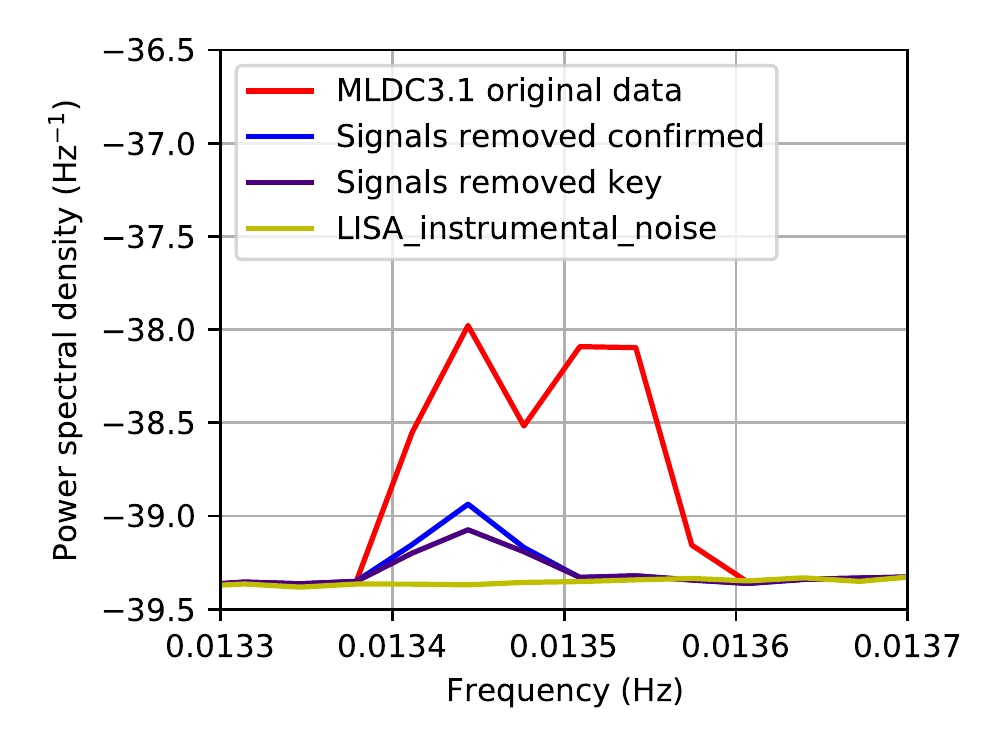}
    \caption{PSDs for different cases.}
    \label{134_one}
\end{figure}

In Fig.~\ref{psd_red_instr}, one may note that in the frequency bands around $0.0135$ Hz and $0.0105$ Hz, the residuals have some high peaks compared with the noise, which indicates that the detection capacity is poor in these frequency bands.
To figure out why the residuals are still high, the injected signals' parameters are used to generate the waveform applying the analytical mathematical formula (i.e., Eq.~\eqref{eq:hp} - \eqref{eq:Phi}), and then subtract them from the original data.
The results are shown in Fig.~\ref{134_one}.
One can find that the residuals are still higher than noise, thus the search method is still trust-worthy.

\section{Conclusions }
\label{sec:conclusion}

The detection of Galaxy DWDs can recover a large amount of information about the galaxy and remove a large amount of foreground signal power to facilitate the search for signals at cosmological distances, such as the GWs of massive black hole binaries at high redshifts.
However, the detection of DWDs faces enormous challenges:
(1) how to effectively confirm signals in a huge number of DWDs ($\sim 10^8$);
(2) how to reduce calculation time while ensuring detection accuracy and quantity.

To solve the two problems, in this paper, we implement the detection process in two steps.
The first step is the coarse search, in which we use the matched filtering method to match the data with the stochastic template bank, and can give a rough estimation of the signals' parameters.
The second step is the fine search, in which we adopt the PSO algorithm to accurately confirm the signal parameters, using the results of the coarse search as a priori.

For coverage of template bank, we have initially setting it as $\eta=0.99$, and this had been proven by a Monte Carlo simulation method.
By removing templates that are too close together and continuing to populate templates randomly until they could no longer be populated, our template bank coverage ends up being very approaching 100\%. 
In this way, the area not covered by the template can be reduced to a minimum,
e.g., Fig.~\ref{Stochastic_template}.

In order to speed up the calculation in coarse search, we have downsampled the data set, undersampled the template, and adopted the FFT algorithm in the calculating of $\mathcal{F}$-statistic.
The combination of these methods allows us to calculate the waveform template about $10^5 \sim 10^6$ per second, 
the calculation of the whole frequency band with a normal core can be done within 24 hr.
After the coarse search, the results of the coarse search are clustered, and then PSO algorithm is used to perform the accurate search within each cluster.
Finally, we confirmed 10\,092 signals, and 8\,600 signals with FF greater than 0.9. 
The order of magnitude of our detected number of sources is consistent with previous studies~\cite{MockLISADataChallengeTaskForce:2007iof, Blaut:2009si, Littenberg:2011zg, Zhang:2021htc}, 
though a different detection method had adopted.


By analyzing the SNRs and PSD of the confirmed signals, error distribution of source parameters, 
and the PSD of residuals, we find that the Galactic confusion noise has a great influence on the detection of DWDs, 
especially at low frequencies.
Our analysis also suggests that our method performs well in searching signals in all frequency bands.

Last but not least, we used a PSO algorithm in the fine search stage.
The search results serves well as a reference, but in terms of efficiency it is not the most efficient search algorithm.
We aim to implement more efficient search algorithm in the fine search stage and build the DWD analysis pipeline for TianQin based on this work in the future.

\begin{acknowledgements}
    This work has been supported by the Guangdong Major Project of Basic and Applied Basic Research (Grant No. 2019B030302001),
    the Natural Science Foundation of China (Grants No. 12173104), 
    the fellowship of China Postdoctoral Science Foundation (Grant No. 2021M703769),
    the Natural Science Foundation of Guangdong Province of China (Grant No. 2022A1515011862).
    We acknowledge the support by National Supercomputer Center in Guangzhou.

\end{acknowledgements}
\appendix
\section{Some parameters of instrument}
\label{sec:instrument_pars}

For LISA, the PSD of the proof-mass noise and the optical-path noise are~\cite{Vallisneri:2004bn}
\begin{equation}
    \begin{aligned}
        S^{\rm acc} &= 2.5 \times 10^{-48} \left( \frac{f}{1\text{Hz}} \right)^{-2} \left[ 1+\left( \frac{10^{-4}\text{Hz}}{f} \right) ^2 \right] \text{Hz}^{-1}, \\
        S^{\rm opt} &= 1.8\times 10^{-37} \left( \frac{f}{1\text{Hz}} \right)^2  \text{Hz}^{-1}.
    \end{aligned}
    \label{eq:noise_acc}
\end{equation}

The arm-length of LISA is $L = 2.5\times 10^6$ km~\cite{LISA:2017pwj},
but one should note that the arm-length that used to generate MLDC-3.1 data is $L = 5\times 10^6$ km.

\section{The Galactic confusion noise}
\label{sec:foreground_noise}

MLDC-3.1 data contains a Galactic GW Galactic confusion from $\sim$ 60 million compact binary systems~\cite{Babak:2008aa}.
When we analysis MLDC-3.1 data we must add an estimate of the confusion noise $S_{n, \rm conf}(f)$ which is derived from data simulations ~\cite{Timpano:2005gm}:
\begin{align}
    & 
    S_{n,\rm conf}(f) = 16(2\pi fL)^{2} \sin^2(2\pi fL) \text{Hz}^{-1} \notag \\
    &
    \times
    \begin{cases}
        10^{-44.62} (f/\text{Hz})^{-2.3}, & f \in \left[10^{-4}, 10^{-3}\right) \text{Hz}, \\
        10^{-50.92} (f/\text{Hz})^{-4.4}, & f \in \left[10^{-3}, 10^{-2.7}\right)\text{Hz}, \\
        10^{-62.8} (f/\text{Hz})^{-8.8}, & f \in \left[10^{-2.7}, 10^{-2.4}\right)\text{Hz}, \\
        10^{-89.68} (f/\text{Hz})^{-20.0}, & f \in\left[10^{-2.4}, 10^{-2.0}\right)\text{Hz}.
    \end{cases}
    \label{eq:confusion} 
\end{align} 

\section{Analytical derivation of the extrinsic parameters}
\label{sec:analytical_Amp}

The $\mathcal{F}$-statistic depends only on $\{\lambda, \beta, f,\dot{f} \}$.
Inverting the four relations $\mathcal{A}^{\mu}(\phi_0, \psi, \iota, h_0 )$ defined in Eq. \eqref{eq:four_amp}, one can obtain the extrinsic parameters analytically.
Firstly, we define two new parameters $A_{\rm sum}$ and $D_a$ as
\begin{align}
    & A_{\rm sum}  \equiv  \sum_{\mu=1}^4 (\mathcal{A}^{\mu})^2 = A_+^2 + A_{\times}^2.
    \label{eq:Asum1} \\
    & D_a  \equiv   \mathcal{A}^{1} \mathcal{A}^{4} - \mathcal{A}^{2} \mathcal{A}^{3} = A_+A_{\times}.
    \label{eq:Da} 
\end{align}
Using Eqs. \eqref{eq:Asum1} and \eqref{eq:Da}, one have
\begin{align}
    & 2A_{+,\times}^2 = A_{\rm sum} \pm \sqrt{A_{\rm sum}^2-4D_a^2}.
    \label{eq:ApAc} 
\end{align}

As defined in Eqs. \eqref{eq:hp} and \eqref{eq:hc}, one have $|A_+| \ge |A_{\times}|$, which means that the sign of  $A_+$ must be positive, while the sign of $A_{\times}$ is determined by the sign of $D_a$.
Thus,
\begin{align}
    A_{+} =& \sqrt{(A_{\rm sum}+\sqrt{A_{\rm sum}^2-4D_a^2})/2}.
    \label{eq:Ap_analy} \\
    A_{\times} =& \sqrt{(A_{\rm sum}-\sqrt{A_{\rm sum}^2-4D_a^2})/2}.
    \label{eq:Ac_analy} 
\end{align}
Combining the above equations, one can easily obtain
\begin{align}
    \psi =& \frac{1}{2}\arctan \left( \frac{\mathcal{A}^4A_+ - \mathcal{A}^1A_{\times}}{\mathcal{A}^3A_+ + \mathcal{A}^2A_{\times}} \right).
    \label{eq:psi_analy} \\
    \phi_0 =& \arctan \left( \frac{\mathcal{A}^3A_+ + \mathcal{A}^2A_{\times}}{\mathcal{A}^4A_{\times} - \mathcal{A}^1A_{+}} \right).
    \label{eq:phi_analy} \\
    h_0 =& A_+ + \sqrt{A_+^2-A_{\times}^2}.
    \label{eq:h0_analy} \\
    \iota =& \arccos\left( \frac{A_{\times}}{h_0} \right).
    \label{eq:iota_analy} 
\end{align}  

\section{Principle of data frequency mixing}
\label{sec:data_shift}

The Fourier Transform (FT) of the cosine function is :
\begin{align}
    & \cos(2\pi f_0 t) \stackrel{FT}{\longleftrightarrow} \pi \left[  \delta(f+f_0)  + \delta(f-f_0)  \right].
    \label{eq:cos_fft} 
\end{align} 
Assuming that $p(t)$ and $q(t)$ are different time domain data, and both have Fourier transforms, namely $\tilde{p}(f)$, $\tilde{q}(f)$. 
Multiplying two data in the time domain is equal to their respective Fourier transform convolution, where ``*'' means convolution,
\begin{align}
    & p(t) q(t)  \stackrel{FT} {\longleftrightarrow} \frac{1}{2\pi} \tilde{p}(f) * \tilde{q}(f).
    \label{eq:twodata_fft} 
\end{align} 
Let $q(t) = p(t)\cos(2\pi f_0 t)$, according to Eq. \eqref{eq:cos_fft} and Eq. \eqref{eq:twodata_fft}  we get:
\begin{equation}
    \begin{aligned}
        \tilde{q}(f) =& \frac{1}{2\pi} \tilde{p}(f) * \pi \left[ \delta(f+f_0)  + \delta(f-f_0) \right] \\
        =& \frac{1}{2}\left[\tilde{p}(f+f_0) +\tilde{p}(f-f_0) \right].
    \end{aligned}
    \label{eq:shif_fft} 
\end{equation} 
Thus, one can find that, we can get two peaks after the data shifting.

\bibliography{refs}

\begin{thebibliography}{97}
\expandafter\ifx\csname natexlab\endcsname\relax\def\natexlab#1{#1}\fi
\expandafter\ifx\csname bibnamefont\endcsname\relax
  \def\bibnamefont#1{#1}\fi
\expandafter\ifx\csname bibfnamefont\endcsname\relax
  \def\bibfnamefont#1{#1}\fi
\expandafter\ifx\csname citenamefont\endcsname\relax
  \def\citenamefont#1{#1}\fi
\expandafter\ifx\csname url\endcsname\relax
  \def\url#1{\texttt{#1}}\fi
\expandafter\ifx\csname urlprefix\endcsname\relax\def\urlprefix{URL }\fi
\providecommand{\bibinfo}[2]{#2}
\providecommand{\eprint}[2][]{\url{#2}}

\bibitem[{\citenamefont{Abbott et~al.}(2016)}]{LIGOScientific:2016aoc}
\bibinfo{author}{\bibfnamefont{B.~P.} \bibnamefont{Abbott}}
  \bibnamefont{et~al.} (\bibinfo{collaboration}{LIGO Scientific, Virgo}),
  \bibinfo{journal}{Phys. Rev. Lett.} \textbf{\bibinfo{volume}{116}},
  \bibinfo{pages}{061102} (\bibinfo{year}{2016}), \eprint{1602.03837}.

\bibitem[{\citenamefont{Abbott et~al.}(2019)}]{LIGOScientific:2018mvr}
\bibinfo{author}{\bibfnamefont{B.~P.} \bibnamefont{Abbott}}
  \bibnamefont{et~al.} (\bibinfo{collaboration}{LIGO Scientific, Virgo}),
  \bibinfo{journal}{Phys. Rev. X} \textbf{\bibinfo{volume}{9}},
  \bibinfo{pages}{031040} (\bibinfo{year}{2019}), \eprint{1811.12907}.

\bibitem[{\citenamefont{Abbott
  et~al.}(2021{\natexlab{a}})}]{LIGOScientific:2020ibl}
\bibinfo{author}{\bibfnamefont{R.}~\bibnamefont{Abbott}} \bibnamefont{et~al.}
  (\bibinfo{collaboration}{LIGO Scientific, Virgo}), \bibinfo{journal}{Phys.
  Rev. X} \textbf{\bibinfo{volume}{11}}, \bibinfo{pages}{021053}
  (\bibinfo{year}{2021}{\natexlab{a}}), \eprint{2010.14527}.

\bibitem[{\citenamefont{Abbott
  et~al.}(2021{\natexlab{b}})}]{LIGOScientific:2021usb}
\bibinfo{author}{\bibfnamefont{R.}~\bibnamefont{Abbott}} \bibnamefont{et~al.}
  (\bibinfo{collaboration}{LIGO Scientific, VIRGO})
  (\bibinfo{year}{2021}{\natexlab{b}}), \eprint{2108.01045}.

\bibitem[{\citenamefont{Abbott
  et~al.}(2021{\natexlab{c}})}]{LIGOScientific:2021djp}
\bibinfo{author}{\bibfnamefont{R.}~\bibnamefont{Abbott}} \bibnamefont{et~al.}
  (\bibinfo{collaboration}{LIGO Scientific, VIRGO, KAGRA})
  (\bibinfo{year}{2021}{\natexlab{c}}), \eprint{2111.03606}.

\bibitem[{\citenamefont{Hall et~al.}(2019)\citenamefont{Hall, Cahillane, Izumi,
  Smith, and Adhikari}}]{Hall:2017off}
\bibinfo{author}{\bibfnamefont{E.~D.} \bibnamefont{Hall}},
  \bibinfo{author}{\bibfnamefont{C.}~\bibnamefont{Cahillane}},
  \bibinfo{author}{\bibfnamefont{K.}~\bibnamefont{Izumi}},
  \bibinfo{author}{\bibfnamefont{R.~J.~E.} \bibnamefont{Smith}},
  \bibnamefont{and} \bibinfo{author}{\bibfnamefont{R.~X.}
  \bibnamefont{Adhikari}}, \bibinfo{journal}{Class. Quant. Grav.}
  \textbf{\bibinfo{volume}{36}}, \bibinfo{pages}{205006}
  (\bibinfo{year}{2019}), \eprint{1712.09719}.

\bibitem[{\citenamefont{Accadia et~al.}(2012)}]{VIRGO:2012dcp}
\bibinfo{author}{\bibfnamefont{T.}~\bibnamefont{Accadia}} \bibnamefont{et~al.}
  (\bibinfo{collaboration}{VIRGO}), \bibinfo{journal}{JINST}
  \textbf{\bibinfo{volume}{7}}, \bibinfo{pages}{P03012} (\bibinfo{year}{2012}).

\bibitem[{\citenamefont{Aso et~al.}(2013)\citenamefont{Aso, Michimura, Somiya,
  Ando, Miyakawa, Sekiguchi, Tatsumi, and Yamamoto}}]{Aso:2013eba}
\bibinfo{author}{\bibfnamefont{Y.}~\bibnamefont{Aso}},
  \bibinfo{author}{\bibfnamefont{Y.}~\bibnamefont{Michimura}},
  \bibinfo{author}{\bibfnamefont{K.}~\bibnamefont{Somiya}},
  \bibinfo{author}{\bibfnamefont{M.}~\bibnamefont{Ando}},
  \bibinfo{author}{\bibfnamefont{O.}~\bibnamefont{Miyakawa}},
  \bibinfo{author}{\bibfnamefont{T.}~\bibnamefont{Sekiguchi}},
  \bibinfo{author}{\bibfnamefont{D.}~\bibnamefont{Tatsumi}}, \bibnamefont{and}
  \bibinfo{author}{\bibfnamefont{H.}~\bibnamefont{Yamamoto}}
  (\bibinfo{collaboration}{KAGRA}), \bibinfo{journal}{Phys. Rev. D}
  \textbf{\bibinfo{volume}{88}}, \bibinfo{pages}{043007}
  (\bibinfo{year}{2013}), \eprint{1306.6747}.

\bibitem[{\citenamefont{Aasi et~al.}(2015)}]{LIGOScientific:2014pky}
\bibinfo{author}{\bibfnamefont{J.}~\bibnamefont{Aasi}} \bibnamefont{et~al.}
  (\bibinfo{collaboration}{LIGO Scientific}), \bibinfo{journal}{Class. Quant.
  Grav.} \textbf{\bibinfo{volume}{32}}, \bibinfo{pages}{074001}
  (\bibinfo{year}{2015}), \eprint{1411.4547}.

\bibitem[{\citenamefont{Acernese et~al.}(2015)}]{VIRGO:2014yos}
\bibinfo{author}{\bibfnamefont{F.}~\bibnamefont{Acernese}} \bibnamefont{et~al.}
  (\bibinfo{collaboration}{VIRGO}), \bibinfo{journal}{Class. Quant. Grav.}
  \textbf{\bibinfo{volume}{32}}, \bibinfo{pages}{024001}
  (\bibinfo{year}{2015}), \eprint{1408.3978}.

\bibitem[{\citenamefont{Somiya}(2012)}]{Somiya:2011np}
\bibinfo{author}{\bibfnamefont{K.}~\bibnamefont{Somiya}}
  (\bibinfo{collaboration}{KAGRA}), \bibinfo{journal}{Class. Quant. Grav.}
  \textbf{\bibinfo{volume}{29}}, \bibinfo{pages}{124007}
  (\bibinfo{year}{2012}), \eprint{1111.7185}.

\bibitem[{\citenamefont{Katz et~al.}(2020)\citenamefont{Katz, Kelley,
  Dosopoulou, Berry, Blecha, and Larson}}]{Katz:2019qlu}
\bibinfo{author}{\bibfnamefont{M.~L.} \bibnamefont{Katz}},
  \bibinfo{author}{\bibfnamefont{L.~Z.} \bibnamefont{Kelley}},
  \bibinfo{author}{\bibfnamefont{F.}~\bibnamefont{Dosopoulou}},
  \bibinfo{author}{\bibfnamefont{S.}~\bibnamefont{Berry}},
  \bibinfo{author}{\bibfnamefont{L.}~\bibnamefont{Blecha}}, \bibnamefont{and}
  \bibinfo{author}{\bibfnamefont{S.~L.} \bibnamefont{Larson}},
  \bibinfo{journal}{Mon. Not. Roy. Astron. Soc.}
  \textbf{\bibinfo{volume}{491}}, \bibinfo{pages}{2301} (\bibinfo{year}{2020}),
  \eprint{1908.05779}.

\bibitem[{\citenamefont{Wang et~al.}(2019)}]{Wang:2019ryf}
\bibinfo{author}{\bibfnamefont{H.-T.} \bibnamefont{Wang}} \bibnamefont{et~al.},
  \bibinfo{journal}{Phys. Rev. D} \textbf{\bibinfo{volume}{100}},
  \bibinfo{pages}{043003} (\bibinfo{year}{2019}), \eprint{1902.04423}.

\bibitem[{\citenamefont{Calabrese et~al.}(2017)}]{Calabrese:2017ypx}
\bibinfo{author}{\bibfnamefont{E.}~\bibnamefont{Calabrese}}
  \bibnamefont{et~al.}, \bibinfo{journal}{Phys. Rev. D}
  \textbf{\bibinfo{volume}{95}}, \bibinfo{pages}{063525}
  (\bibinfo{year}{2017}), \eprint{1702.03272}.

\bibitem[{\citenamefont{Fan et~al.}(2020)\citenamefont{Fan, Hu, Barausse,
  Sesana, Zhang, Zhang, Zi, and Mei}}]{Fan:2020zhy}
\bibinfo{author}{\bibfnamefont{H.-M.} \bibnamefont{Fan}},
  \bibinfo{author}{\bibfnamefont{Y.-M.} \bibnamefont{Hu}},
  \bibinfo{author}{\bibfnamefont{E.}~\bibnamefont{Barausse}},
  \bibinfo{author}{\bibfnamefont{A.}~\bibnamefont{Sesana}},
  \bibinfo{author}{\bibfnamefont{J.-d.} \bibnamefont{Zhang}},
  \bibinfo{author}{\bibfnamefont{X.}~\bibnamefont{Zhang}},
  \bibinfo{author}{\bibfnamefont{T.-G.} \bibnamefont{Zi}}, \bibnamefont{and}
  \bibinfo{author}{\bibfnamefont{J.}~\bibnamefont{Mei}},
  \bibinfo{journal}{Phys. Rev. D} \textbf{\bibinfo{volume}{102}},
  \bibinfo{pages}{063016} (\bibinfo{year}{2020}), \eprint{2005.08212}.

\bibitem[{\citenamefont{Sesana}(2016)}]{Sesana:2016ljz}
\bibinfo{author}{\bibfnamefont{A.}~\bibnamefont{Sesana}},
  \bibinfo{journal}{Phys. Rev. Lett.} \textbf{\bibinfo{volume}{116}},
  \bibinfo{pages}{231102} (\bibinfo{year}{2016}), \eprint{1602.06951}.

\bibitem[{\citenamefont{Kyutoku and Seto}(2016)}]{Kyutoku:2016ppx}
\bibinfo{author}{\bibfnamefont{K.}~\bibnamefont{Kyutoku}} \bibnamefont{and}
  \bibinfo{author}{\bibfnamefont{N.}~\bibnamefont{Seto}},
  \bibinfo{journal}{Mon. Not. Roy. Astron. Soc.}
  \textbf{\bibinfo{volume}{462}}, \bibinfo{pages}{2177} (\bibinfo{year}{2016}),
  \eprint{1606.02298}.

\bibitem[{\citenamefont{Liu et~al.}(2020)\citenamefont{Liu, Hu, Zhang, and
  Mei}}]{Liu:2020eko}
\bibinfo{author}{\bibfnamefont{S.}~\bibnamefont{Liu}},
  \bibinfo{author}{\bibfnamefont{Y.-M.} \bibnamefont{Hu}},
  \bibinfo{author}{\bibfnamefont{J.-d.} \bibnamefont{Zhang}}, \bibnamefont{and}
  \bibinfo{author}{\bibfnamefont{J.}~\bibnamefont{Mei}},
  \bibinfo{journal}{Phys. Rev. D} \textbf{\bibinfo{volume}{101}},
  \bibinfo{pages}{103027} (\bibinfo{year}{2020}), \eprint{2004.14242}.

\bibitem[{\citenamefont{Nelemans
  et~al.}(2001{\natexlab{a}})\citenamefont{Nelemans, Yungelson, and
  Portegies~Zwart}}]{Nelemans:2001hp}
\bibinfo{author}{\bibfnamefont{G.}~\bibnamefont{Nelemans}},
  \bibinfo{author}{\bibfnamefont{L.~R.} \bibnamefont{Yungelson}},
  \bibnamefont{and} \bibinfo{author}{\bibfnamefont{S.~F.}
  \bibnamefont{Portegies~Zwart}}, \bibinfo{journal}{Astron. Astrophys.}
  \textbf{\bibinfo{volume}{375}}, \bibinfo{pages}{890}
  (\bibinfo{year}{2001}{\natexlab{a}}), \eprint{astro-ph/0105221}.

\bibitem[{\citenamefont{Yu and Jeffery}(2010)}]{Yu:2010fq}
\bibinfo{author}{\bibfnamefont{S.}~\bibnamefont{Yu}} \bibnamefont{and}
  \bibinfo{author}{\bibfnamefont{C.~S.} \bibnamefont{Jeffery}},
  \bibinfo{journal}{Astron. Astrophys.} \textbf{\bibinfo{volume}{521}},
  \bibinfo{pages}{A85} (\bibinfo{year}{2010}), \eprint{1007.4267}.

\bibitem[{\citenamefont{Breivik et~al.}(2020)}]{Breivik:2019lmt}
\bibinfo{author}{\bibfnamefont{K.}~\bibnamefont{Breivik}} \bibnamefont{et~al.},
  \bibinfo{journal}{Astrophys. J.} \textbf{\bibinfo{volume}{898}},
  \bibinfo{pages}{71} (\bibinfo{year}{2020}), \eprint{1911.00903}.

\bibitem[{\citenamefont{Huang et~al.}(2020)\citenamefont{Huang, Hu, Korol, Li,
  Liang, Lu, Wang, Yu, and Mei}}]{Huang:2020rjf}
\bibinfo{author}{\bibfnamefont{S.-J.} \bibnamefont{Huang}},
  \bibinfo{author}{\bibfnamefont{Y.-M.} \bibnamefont{Hu}},
  \bibinfo{author}{\bibfnamefont{V.}~\bibnamefont{Korol}},
  \bibinfo{author}{\bibfnamefont{P.-C.} \bibnamefont{Li}},
  \bibinfo{author}{\bibfnamefont{Z.-C.} \bibnamefont{Liang}},
  \bibinfo{author}{\bibfnamefont{Y.}~\bibnamefont{Lu}},
  \bibinfo{author}{\bibfnamefont{H.-T.} \bibnamefont{Wang}},
  \bibinfo{author}{\bibfnamefont{S.}~\bibnamefont{Yu}}, \bibnamefont{and}
  \bibinfo{author}{\bibfnamefont{J.}~\bibnamefont{Mei}},
  \bibinfo{journal}{Phys. Rev. D} \textbf{\bibinfo{volume}{102}},
  \bibinfo{pages}{063021} (\bibinfo{year}{2020}), \eprint{2005.07889}.

\bibitem[{\citenamefont{{Romano} and {Cornish}}(2017)}]{2017LRR....20....2R}
\bibinfo{author}{\bibfnamefont{J.~D.} \bibnamefont{{Romano}}} \bibnamefont{and}
  \bibinfo{author}{\bibfnamefont{N.~J.} \bibnamefont{{Cornish}}},
  \bibinfo{journal}{Living Reviews in Relativity}
  \textbf{\bibinfo{volume}{20}}, \bibinfo{eid}{2} (\bibinfo{year}{2017}),
  \eprint{1608.06889}.

\bibitem[{\citenamefont{Liang et~al.}(2021)\citenamefont{Liang, Hu, Jiang,
  Cheng, Zhang, and Mei}}]{Liang:2021bde}
\bibinfo{author}{\bibfnamefont{Z.-C.} \bibnamefont{Liang}},
  \bibinfo{author}{\bibfnamefont{Y.-M.} \bibnamefont{Hu}},
  \bibinfo{author}{\bibfnamefont{Y.}~\bibnamefont{Jiang}},
  \bibinfo{author}{\bibfnamefont{J.}~\bibnamefont{Cheng}},
  \bibinfo{author}{\bibfnamefont{J.-d.} \bibnamefont{Zhang}}, \bibnamefont{and}
  \bibinfo{author}{\bibfnamefont{J.}~\bibnamefont{Mei}} (\bibinfo{year}{2021}),
  \eprint{2107.08643}.

\bibitem[{\citenamefont{Luo et~al.}(2016)}]{TianQin:2015yph}
\bibinfo{author}{\bibfnamefont{J.}~\bibnamefont{Luo}} \bibnamefont{et~al.}
  (\bibinfo{collaboration}{TianQin}), \bibinfo{journal}{Class. Quant. Grav.}
  \textbf{\bibinfo{volume}{33}}, \bibinfo{pages}{035010}
  (\bibinfo{year}{2016}), \eprint{1512.02076}.

\bibitem[{\citenamefont{Amaro-Seoane et~al.}(2017)}]{LISA:2017pwj}
\bibinfo{author}{\bibfnamefont{P.}~\bibnamefont{Amaro-Seoane}}
  \bibnamefont{et~al.} (\bibinfo{collaboration}{LISA}) (\bibinfo{year}{2017}),
  \eprint{1702.00786}.

\bibitem[{\citenamefont{Lamberts et~al.}(2018)\citenamefont{Lamberts,
  Garrison-Kimmel, Hopkins, Quataert, Bullock, Faucher-Gigu\`ere, Wetzel,
  Keres, Drango, and Sanderson}}]{Lamberts:2018cge}
\bibinfo{author}{\bibfnamefont{A.}~\bibnamefont{Lamberts}},
  \bibinfo{author}{\bibfnamefont{S.}~\bibnamefont{Garrison-Kimmel}},
  \bibinfo{author}{\bibfnamefont{P.}~\bibnamefont{Hopkins}},
  \bibinfo{author}{\bibfnamefont{E.}~\bibnamefont{Quataert}},
  \bibinfo{author}{\bibfnamefont{J.}~\bibnamefont{Bullock}},
  \bibinfo{author}{\bibfnamefont{C.-A.} \bibnamefont{Faucher-Gigu\`ere}},
  \bibinfo{author}{\bibfnamefont{A.}~\bibnamefont{Wetzel}},
  \bibinfo{author}{\bibfnamefont{D.}~\bibnamefont{Keres}},
  \bibinfo{author}{\bibfnamefont{K.}~\bibnamefont{Drango}}, \bibnamefont{and}
  \bibinfo{author}{\bibfnamefont{R.}~\bibnamefont{Sanderson}},
  \bibinfo{journal}{Mon. Not. Roy. Astron. Soc.}
  \textbf{\bibinfo{volume}{480}}, \bibinfo{pages}{2704} (\bibinfo{year}{2018}),
  \eprint{1801.03099}.

\bibitem[{\citenamefont{Barone-Nugent et~al.}(2012)}]{Barone-Nugent:2012esy}
\bibinfo{author}{\bibfnamefont{R.~L.} \bibnamefont{Barone-Nugent}}
  \bibnamefont{et~al.}, \bibinfo{journal}{Mon. Not. Roy. Astron. Soc.}
  \textbf{\bibinfo{volume}{425}}, \bibinfo{pages}{1007} (\bibinfo{year}{2012}),
  \eprint{1204.2308}.

\bibitem[{\citenamefont{Perlmutter
  et~al.}(1999)}]{SupernovaCosmologyProject:1998vns}
\bibinfo{author}{\bibfnamefont{S.}~\bibnamefont{Perlmutter}}
  \bibnamefont{et~al.} (\bibinfo{collaboration}{Supernova Cosmology Project}),
  \bibinfo{journal}{Astrophys. J.} \textbf{\bibinfo{volume}{517}},
  \bibinfo{pages}{565} (\bibinfo{year}{1999}), \eprint{astro-ph/9812133}.

\bibitem[{\citenamefont{Riess et~al.}(1998)}]{SupernovaSearchTeam:1998fmf}
\bibinfo{author}{\bibfnamefont{A.~G.} \bibnamefont{Riess}} \bibnamefont{et~al.}
  (\bibinfo{collaboration}{Supernova Search Team}), \bibinfo{journal}{Astron.
  J.} \textbf{\bibinfo{volume}{116}}, \bibinfo{pages}{1009}
  (\bibinfo{year}{1998}), \eprint{astro-ph/9805201}.

\bibitem[{\citenamefont{{Postnov} and {Yungelson}}(2014)}]{2014LRR....17....3P}
\bibinfo{author}{\bibfnamefont{K.~A.} \bibnamefont{{Postnov}}}
  \bibnamefont{and} \bibinfo{author}{\bibfnamefont{L.~R.}
  \bibnamefont{{Yungelson}}}, \bibinfo{journal}{Living Reviews in Relativity}
  \textbf{\bibinfo{volume}{17}}, \bibinfo{eid}{3} (\bibinfo{year}{2014}),
  \eprint{1403.4754}.

\bibitem[{\citenamefont{Belczynski et~al.}(2001)\citenamefont{Belczynski,
  Kalogera, and Bulik}}]{Belczynski:2001uc}
\bibinfo{author}{\bibfnamefont{K.}~\bibnamefont{Belczynski}},
  \bibinfo{author}{\bibfnamefont{V.}~\bibnamefont{Kalogera}}, \bibnamefont{and}
  \bibinfo{author}{\bibfnamefont{T.}~\bibnamefont{Bulik}},
  \bibinfo{journal}{Astrophys. J.} \textbf{\bibinfo{volume}{572}},
  \bibinfo{pages}{407} (\bibinfo{year}{2001}), \eprint{astro-ph/0111452}.

\bibitem[{\citenamefont{Nelemans
  et~al.}(2001{\natexlab{b}})\citenamefont{Nelemans, Yungelson,
  Portegies~Zwart, and Verbunt}}]{Nelemans:2000es}
\bibinfo{author}{\bibfnamefont{G.}~\bibnamefont{Nelemans}},
  \bibinfo{author}{\bibfnamefont{L.~R.} \bibnamefont{Yungelson}},
  \bibinfo{author}{\bibfnamefont{S.~F.} \bibnamefont{Portegies~Zwart}},
  \bibnamefont{and} \bibinfo{author}{\bibfnamefont{F.}~\bibnamefont{Verbunt}},
  \bibinfo{journal}{Astron. Astrophys.} \textbf{\bibinfo{volume}{365}},
  \bibinfo{pages}{491} (\bibinfo{year}{2001}{\natexlab{b}}),
  \eprint{astro-ph/0010457}.

\bibitem[{\citenamefont{Marsh et~al.}(2004)\citenamefont{Marsh, Nelemans, and
  Steeghs}}]{Marsh:2003rd}
\bibinfo{author}{\bibfnamefont{T.~R.} \bibnamefont{Marsh}},
  \bibinfo{author}{\bibfnamefont{G.}~\bibnamefont{Nelemans}}, \bibnamefont{and}
  \bibinfo{author}{\bibfnamefont{D.}~\bibnamefont{Steeghs}},
  \bibinfo{journal}{Mon. Not. Roy. Astron. Soc.}
  \textbf{\bibinfo{volume}{350}}, \bibinfo{pages}{113} (\bibinfo{year}{2004}),
  \eprint{astro-ph/0312577}.

\bibitem[{\citenamefont{{Solheim}}(2010)}]{2010PASP..122.1133S}
\bibinfo{author}{\bibfnamefont{J.~E.} \bibnamefont{{Solheim}}},
  \bibinfo{journal}{Publications of the ASP} \textbf{\bibinfo{volume}{122}},
  \bibinfo{pages}{1133} (\bibinfo{year}{2010}).

\bibitem[{\citenamefont{Tauris}(2018)}]{Tauris:2018kzq}
\bibinfo{author}{\bibfnamefont{T.~M.} \bibnamefont{Tauris}},
  \bibinfo{journal}{Phys. Rev. Lett.} \textbf{\bibinfo{volume}{121}},
  \bibinfo{pages}{131105} (\bibinfo{year}{2018}), \bibinfo{note}{[Erratum:
  Phys.Rev.Lett. 124, 149902 (2020)]}, \eprint{1809.03504}.

\bibitem[{\citenamefont{Fuller and Lai}(2012)}]{Fuller:2012ky}
\bibinfo{author}{\bibfnamefont{J.}~\bibnamefont{Fuller}} \bibnamefont{and}
  \bibinfo{author}{\bibfnamefont{D.}~\bibnamefont{Lai}},
  \bibinfo{journal}{Astrophys. J. Lett.} \textbf{\bibinfo{volume}{756}},
  \bibinfo{pages}{17} (\bibinfo{year}{2012}), \eprint{1206.0470}.

\bibitem[{\citenamefont{Dall'Osso and Rossi}(2014)}]{DallOsso:2013uac}
\bibinfo{author}{\bibfnamefont{S.}~\bibnamefont{Dall'Osso}} \bibnamefont{and}
  \bibinfo{author}{\bibfnamefont{E.~M.} \bibnamefont{Rossi}},
  \bibinfo{journal}{Mon. Not. Roy. Astron. Soc.}
  \textbf{\bibinfo{volume}{443}}, \bibinfo{pages}{1057} (\bibinfo{year}{2014}),
  \eprint{1308.1664}.

\bibitem[{\citenamefont{Benacquista and
  Holley-Bockelmann}(2006)}]{Benacquista:2005tm}
\bibinfo{author}{\bibfnamefont{M.}~\bibnamefont{Benacquista}} \bibnamefont{and}
  \bibinfo{author}{\bibfnamefont{K.}~\bibnamefont{Holley-Bockelmann}},
  \bibinfo{journal}{Astrophys. J.} \textbf{\bibinfo{volume}{645}},
  \bibinfo{pages}{589} (\bibinfo{year}{2006}), \eprint{astro-ph/0504135}.

\bibitem[{\citenamefont{Adams et~al.}(2012)\citenamefont{Adams, Cornish, and
  Littenberg}}]{Adams:2012qw}
\bibinfo{author}{\bibfnamefont{M.~R.} \bibnamefont{Adams}},
  \bibinfo{author}{\bibfnamefont{N.~J.} \bibnamefont{Cornish}},
  \bibnamefont{and} \bibinfo{author}{\bibfnamefont{T.~B.}
  \bibnamefont{Littenberg}}, \bibinfo{journal}{Phys. Rev. D}
  \textbf{\bibinfo{volume}{86}}, \bibinfo{pages}{124032}
  (\bibinfo{year}{2012}), \eprint{1209.6286}.

\bibitem[{\citenamefont{Korol et~al.}(2019)\citenamefont{Korol, Rossi, and
  Barausse}}]{Korol:2018wep}
\bibinfo{author}{\bibfnamefont{V.}~\bibnamefont{Korol}},
  \bibinfo{author}{\bibfnamefont{E.~M.} \bibnamefont{Rossi}}, \bibnamefont{and}
  \bibinfo{author}{\bibfnamefont{E.}~\bibnamefont{Barausse}},
  \bibinfo{journal}{Mon. Not. Roy. Astron. Soc.}
  \textbf{\bibinfo{volume}{483}}, \bibinfo{pages}{5518} (\bibinfo{year}{2019}),
  \eprint{1806.03306}.

\bibitem[{\citenamefont{Wilhelm et~al.}(2020)\citenamefont{Wilhelm, Korol,
  Rossi, and D'Onghia}}]{Wilhelm:2020qjc}
\bibinfo{author}{\bibfnamefont{M.~J.~C.} \bibnamefont{Wilhelm}},
  \bibinfo{author}{\bibfnamefont{V.}~\bibnamefont{Korol}},
  \bibinfo{author}{\bibfnamefont{E.~M.} \bibnamefont{Rossi}}, \bibnamefont{and}
  \bibinfo{author}{\bibfnamefont{E.}~\bibnamefont{D'Onghia}},
  \bibinfo{journal}{Mon. Not. Roy. Astron. Soc.}
  \textbf{\bibinfo{volume}{500}}, \bibinfo{pages}{4958} (\bibinfo{year}{2020}),
  \eprint{2003.11074}.

\bibitem[{\citenamefont{Timpano et~al.}(2006)\citenamefont{Timpano, Rubbo, and
  Cornish}}]{Timpano:2005gm}
\bibinfo{author}{\bibfnamefont{S.~E.} \bibnamefont{Timpano}},
  \bibinfo{author}{\bibfnamefont{L.~J.} \bibnamefont{Rubbo}}, \bibnamefont{and}
  \bibinfo{author}{\bibfnamefont{N.~J.} \bibnamefont{Cornish}},
  \bibinfo{journal}{Phys. Rev. D} \textbf{\bibinfo{volume}{73}},
  \bibinfo{pages}{122001} (\bibinfo{year}{2006}), \eprint{gr-qc/0504071}.

\bibitem[{\citenamefont{Crowder and Cornish}(2004)}]{Crowder:2004ca}
\bibinfo{author}{\bibfnamefont{J.}~\bibnamefont{Crowder}} \bibnamefont{and}
  \bibinfo{author}{\bibfnamefont{N.~J.} \bibnamefont{Cornish}},
  \bibinfo{journal}{Phys. Rev. D} \textbf{\bibinfo{volume}{70}},
  \bibinfo{pages}{082004} (\bibinfo{year}{2004}), \eprint{gr-qc/0404129}.

\bibitem[{\citenamefont{Babak et~al.}(2008{\natexlab{a}})\citenamefont{Babak,
  Baker, Benacquista, Cornish, Crowder, Cutler, Larson, Littenberg, Porter,
  Vallisneri et~al.}}]{MockLISADataChallengeTaskForce:2007iof}
\bibinfo{author}{\bibfnamefont{S.}~\bibnamefont{Babak}},
  \bibinfo{author}{\bibfnamefont{J.~G.} \bibnamefont{Baker}},
  \bibinfo{author}{\bibfnamefont{M.~J.} \bibnamefont{Benacquista}},
  \bibinfo{author}{\bibfnamefont{N.~J.} \bibnamefont{Cornish}},
  \bibinfo{author}{\bibfnamefont{J.}~\bibnamefont{Crowder}},
  \bibinfo{author}{\bibfnamefont{C.}~\bibnamefont{Cutler}},
  \bibinfo{author}{\bibfnamefont{S.~L.} \bibnamefont{Larson}},
  \bibinfo{author}{\bibfnamefont{T.~B.} \bibnamefont{Littenberg}},
  \bibinfo{author}{\bibfnamefont{E.~K.} \bibnamefont{Porter}},
  \bibinfo{author}{\bibfnamefont{M.}~\bibnamefont{Vallisneri}},
  \bibnamefont{et~al.} (\bibinfo{collaboration}{Mock LISA Data Challenge Task
  Force}), \bibinfo{journal}{Class. Quant. Grav.}
  \textbf{\bibinfo{volume}{25}}, \bibinfo{pages}{114037}
  (\bibinfo{year}{2008}{\natexlab{a}}), \eprint{0711.2667}.

\bibitem[{\citenamefont{Babak et~al.}(2010)\citenamefont{Babak, Baker,
  Benacquista, Cornish, Larson, Mandel, McWilliams, Petiteau, Porter, Robinson
  et~al.}}]{MockLISADataChallengeTaskForce:2009wir}
\bibinfo{author}{\bibfnamefont{S.}~\bibnamefont{Babak}},
  \bibinfo{author}{\bibfnamefont{J.~G.} \bibnamefont{Baker}},
  \bibinfo{author}{\bibfnamefont{M.~J.} \bibnamefont{Benacquista}},
  \bibinfo{author}{\bibfnamefont{N.~J.} \bibnamefont{Cornish}},
  \bibinfo{author}{\bibfnamefont{S.~L.} \bibnamefont{Larson}},
  \bibinfo{author}{\bibfnamefont{I.}~\bibnamefont{Mandel}},
  \bibinfo{author}{\bibfnamefont{S.~T.} \bibnamefont{McWilliams}},
  \bibinfo{author}{\bibfnamefont{A.}~\bibnamefont{Petiteau}},
  \bibinfo{author}{\bibfnamefont{E.~K.} \bibnamefont{Porter}},
  \bibinfo{author}{\bibfnamefont{E.~L.} \bibnamefont{Robinson}},
  \bibnamefont{et~al.} (\bibinfo{collaboration}{Mock LISA Data Challenge Task
  Force}), \bibinfo{journal}{Class. Quant. Grav.}
  \textbf{\bibinfo{volume}{27}}, \bibinfo{pages}{084009}
  (\bibinfo{year}{2010}), \eprint{0912.0548}.

\bibitem[{\citenamefont{Babak et~al.}(2008{\natexlab{b}})\citenamefont{Babak,
  Baker, Benacquista, Cornish, Crowder, Larson, Plagnol, Porter, Vallisneri,
  Vecchio et~al.}}]{Babak:2008aa}
\bibinfo{author}{\bibfnamefont{S.}~\bibnamefont{Babak}},
  \bibinfo{author}{\bibfnamefont{J.~G.} \bibnamefont{Baker}},
  \bibinfo{author}{\bibfnamefont{M.~J.} \bibnamefont{Benacquista}},
  \bibinfo{author}{\bibfnamefont{N.~J.} \bibnamefont{Cornish}},
  \bibinfo{author}{\bibfnamefont{J.}~\bibnamefont{Crowder}},
  \bibinfo{author}{\bibfnamefont{S.~L.} \bibnamefont{Larson}},
  \bibinfo{author}{\bibfnamefont{E.}~\bibnamefont{Plagnol}},
  \bibinfo{author}{\bibfnamefont{E.~K.} \bibnamefont{Porter}},
  \bibinfo{author}{\bibfnamefont{M.}~\bibnamefont{Vallisneri}},
  \bibinfo{author}{\bibfnamefont{A.}~\bibnamefont{Vecchio}},
  \bibnamefont{et~al.}, \bibinfo{journal}{Class. Quant. Grav.}
  \textbf{\bibinfo{volume}{25}}, \bibinfo{pages}{184026}
  (\bibinfo{year}{2008}{\natexlab{b}}), \eprint{0806.2110}.

\bibitem[{\citenamefont{Arnaud et~al.}(2007)\citenamefont{Arnaud, Auger, Babak,
  Baker, Benacquista, Bloomer, Brown, Camp, Cannizzo, Christensen
  et~al.}}]{Arnaud:2007vr}
\bibinfo{author}{\bibfnamefont{K.~A.} \bibnamefont{Arnaud}},
  \bibinfo{author}{\bibfnamefont{G.}~\bibnamefont{Auger}},
  \bibinfo{author}{\bibfnamefont{S.}~\bibnamefont{Babak}},
  \bibinfo{author}{\bibfnamefont{J.~G.} \bibnamefont{Baker}},
  \bibinfo{author}{\bibfnamefont{M.~J.} \bibnamefont{Benacquista}},
  \bibinfo{author}{\bibfnamefont{E.}~\bibnamefont{Bloomer}},
  \bibinfo{author}{\bibfnamefont{D.~A.} \bibnamefont{Brown}},
  \bibinfo{author}{\bibfnamefont{J.~B.} \bibnamefont{Camp}},
  \bibinfo{author}{\bibfnamefont{J.~K.} \bibnamefont{Cannizzo}},
  \bibinfo{author}{\bibfnamefont{N.}~\bibnamefont{Christensen}},
  \bibnamefont{et~al.}, \bibinfo{journal}{Class. Quant. Grav.}
  \textbf{\bibinfo{volume}{24}}, \bibinfo{pages}{S529} (\bibinfo{year}{2007}),
  \eprint{gr-qc/0701139}.

\bibitem[{\citenamefont{{Arnaud} et~al.}(2007)\citenamefont{{Arnaud}, {Babak},
  {Baker}, {Benacquista}, {Cornish}, {Cutler}, {Finn}, {Larson}, {Littenberg},
  {Porter} et~al.}}]{2007CQGra..24S.551A}
\bibinfo{author}{\bibfnamefont{K.~A.} \bibnamefont{{Arnaud}}},
  \bibinfo{author}{\bibfnamefont{S.}~\bibnamefont{{Babak}}},
  \bibinfo{author}{\bibfnamefont{J.~G.} \bibnamefont{{Baker}}},
  \bibinfo{author}{\bibfnamefont{M.~J.} \bibnamefont{{Benacquista}}},
  \bibinfo{author}{\bibfnamefont{N.~J.} \bibnamefont{{Cornish}}},
  \bibinfo{author}{\bibfnamefont{C.}~\bibnamefont{{Cutler}}},
  \bibinfo{author}{\bibfnamefont{L.~S.} \bibnamefont{{Finn}}},
  \bibinfo{author}{\bibfnamefont{S.~L.} \bibnamefont{{Larson}}},
  \bibinfo{author}{\bibfnamefont{T.}~\bibnamefont{{Littenberg}}},
  \bibinfo{author}{\bibfnamefont{E.~K.} \bibnamefont{{Porter}}},
  \bibnamefont{et~al.}, \bibinfo{journal}{Classical and Quantum Gravity}
  \textbf{\bibinfo{volume}{24}}, \bibinfo{pages}{S551} (\bibinfo{year}{2007}),
  \eprint{gr-qc/0701170}.

\bibitem[{\citenamefont{Arnaud et~al.}(2006)}]{Arnaud:2006gm}
\bibinfo{author}{\bibfnamefont{K.~A.} \bibnamefont{Arnaud}}
  \bibnamefont{et~al.}, \bibinfo{journal}{AIP Conf. Proc.}
  \textbf{\bibinfo{volume}{873}}, \bibinfo{pages}{619} (\bibinfo{year}{2006}),
  \eprint{gr-qc/0609105}.

\bibitem[{\citenamefont{Helstr\"{o}m}(1968)}]{helstrom1968international}
\bibinfo{author}{\bibfnamefont{C.}~\bibnamefont{Helstr\"{o}m}},
  \emph{\bibinfo{title}{International series of monographs in electronics and
  instrumentation, vol. 9, statistical theory of signal detection}}
  (\bibinfo{year}{1968}).

\bibitem[{\citenamefont{B{\l}aut et~al.}(2010)\citenamefont{B{\l}aut, Babak,
  and Kr{\'o}lak}}]{Blaut:2009si}
\bibinfo{author}{\bibfnamefont{A.}~\bibnamefont{B{\l}aut}},
  \bibinfo{author}{\bibfnamefont{S.}~\bibnamefont{Babak}}, \bibnamefont{and}
  \bibinfo{author}{\bibfnamefont{A.}~\bibnamefont{Kr{\'o}lak}},
  \bibinfo{journal}{Phys. Rev. D} \textbf{\bibinfo{volume}{81}},
  \bibinfo{pages}{063008} (\bibinfo{year}{2010}), \eprint{0911.3020}.

\bibitem[{\citenamefont{Cornish and Crowder}(2005)}]{Cornish:2005qw}
\bibinfo{author}{\bibfnamefont{N.~J.} \bibnamefont{Cornish}} \bibnamefont{and}
  \bibinfo{author}{\bibfnamefont{J.}~\bibnamefont{Crowder}},
  \bibinfo{journal}{Phys. Rev. D} \textbf{\bibinfo{volume}{72}},
  \bibinfo{pages}{043005} (\bibinfo{year}{2005}), \eprint{gr-qc/0506059}.

\bibitem[{\citenamefont{Crowder and Cornish}(2007)}]{Crowder:2007ft}
\bibinfo{author}{\bibfnamefont{J.}~\bibnamefont{Crowder}} \bibnamefont{and}
  \bibinfo{author}{\bibfnamefont{N.~J.} \bibnamefont{Cornish}},
  \bibinfo{journal}{Class. Quant. Grav.} \textbf{\bibinfo{volume}{24}},
  \bibinfo{pages}{S575} (\bibinfo{year}{2007}), \eprint{0704.2917}.

\bibitem[{\citenamefont{Littenberg}(2011)}]{Littenberg:2011zg}
\bibinfo{author}{\bibfnamefont{T.~B.} \bibnamefont{Littenberg}},
  \bibinfo{journal}{Phys. Rev. D} \textbf{\bibinfo{volume}{84}},
  \bibinfo{pages}{063009} (\bibinfo{year}{2011}), \eprint{1106.6355}.

\bibitem[{\citenamefont{Zhang et~al.}(2021)\citenamefont{Zhang, Mohanty, Zou,
  and Liu}}]{Zhang:2021htc}
\bibinfo{author}{\bibfnamefont{X.}~\bibnamefont{Zhang}},
  \bibinfo{author}{\bibfnamefont{S.~D.} \bibnamefont{Mohanty}},
  \bibinfo{author}{\bibfnamefont{X.}~\bibnamefont{Zou}}, \bibnamefont{and}
  \bibinfo{author}{\bibfnamefont{Y.}~\bibnamefont{Liu}},
  \bibinfo{journal}{Phys. Rev. D} \textbf{\bibinfo{volume}{104}},
  \bibinfo{pages}{024023} (\bibinfo{year}{2021}), \eprint{2103.09391}.

\bibitem[{\citenamefont{{John Conway and Neil J. A. Sloane}}(1999)}]{RN262}
\bibinfo{author}{\bibnamefont{{John Conway and Neil J. A. Sloane}}},
  \emph{\bibinfo{title}{Sphere Packings, Lattices and Groups}}
  (\bibinfo{publisher}{Springer-Verlag}, \bibinfo{address}{New York},
  \bibinfo{year}{1999}), \bibinfo{edition}{3rd} ed., ISBN
  \bibinfo{isbn}{978-0-387-98585-5}.

\bibitem[{\citenamefont{Prix}(2007)}]{Prix:2006wm}
\bibinfo{author}{\bibfnamefont{R.}~\bibnamefont{Prix}}, \bibinfo{journal}{Phys.
  Rev. D} \textbf{\bibinfo{volume}{75}}, \bibinfo{pages}{023004}
  (\bibinfo{year}{2007}), \bibinfo{note}{[Erratum: Phys.Rev.D 75, 069901
  (2007)]}, \eprint{gr-qc/0606088}.

\bibitem[{\citenamefont{Messenger et~al.}(2009)\citenamefont{Messenger, Prix,
  and Papa}}]{Messenger:2008ta}
\bibinfo{author}{\bibfnamefont{C.}~\bibnamefont{Messenger}},
  \bibinfo{author}{\bibfnamefont{R.}~\bibnamefont{Prix}}, \bibnamefont{and}
  \bibinfo{author}{\bibfnamefont{M.~A.} \bibnamefont{Papa}},
  \bibinfo{journal}{Phys. Rev. D} \textbf{\bibinfo{volume}{79}},
  \bibinfo{pages}{104017} (\bibinfo{year}{2009}), \eprint{0809.5223}.

\bibitem[{\citenamefont{Astone et~al.}(2010)\citenamefont{Astone, Borkowski,
  Jaranowski, Pietka, and Krolak}}]{Astone:2010ct}
\bibinfo{author}{\bibfnamefont{P.}~\bibnamefont{Astone}},
  \bibinfo{author}{\bibfnamefont{K.~M.} \bibnamefont{Borkowski}},
  \bibinfo{author}{\bibfnamefont{P.}~\bibnamefont{Jaranowski}},
  \bibinfo{author}{\bibfnamefont{M.}~\bibnamefont{Pietka}}, \bibnamefont{and}
  \bibinfo{author}{\bibfnamefont{A.}~\bibnamefont{Krolak}},
  \bibinfo{journal}{Phys. Rev. D} \textbf{\bibinfo{volume}{82}},
  \bibinfo{pages}{022005} (\bibinfo{year}{2010}), \eprint{1003.0844}.

\bibitem[{\citenamefont{Allen et~al.}(2012)\citenamefont{Allen, Anderson,
  Brady, Brown, and Creighton}}]{Allen:2005fk}
\bibinfo{author}{\bibfnamefont{B.}~\bibnamefont{Allen}},
  \bibinfo{author}{\bibfnamefont{W.~G.} \bibnamefont{Anderson}},
  \bibinfo{author}{\bibfnamefont{P.~R.} \bibnamefont{Brady}},
  \bibinfo{author}{\bibfnamefont{D.~A.} \bibnamefont{Brown}}, \bibnamefont{and}
  \bibinfo{author}{\bibfnamefont{J.~D.~E.} \bibnamefont{Creighton}},
  \bibinfo{journal}{Phys. Rev. D} \textbf{\bibinfo{volume}{85}},
  \bibinfo{pages}{122006} (\bibinfo{year}{2012}), \eprint{gr-qc/0509116}.

\bibitem[{\citenamefont{Babak et~al.}(2006)\citenamefont{Babak,
  Balasubramanian, Churches, Cokelaer, and Sathyaprakash}}]{Babak:2006ty}
\bibinfo{author}{\bibfnamefont{S.}~\bibnamefont{Babak}},
  \bibinfo{author}{\bibfnamefont{R.}~\bibnamefont{Balasubramanian}},
  \bibinfo{author}{\bibfnamefont{D.}~\bibnamefont{Churches}},
  \bibinfo{author}{\bibfnamefont{T.}~\bibnamefont{Cokelaer}}, \bibnamefont{and}
  \bibinfo{author}{\bibfnamefont{B.~S.} \bibnamefont{Sathyaprakash}},
  \bibinfo{journal}{Class. Quant. Grav.} \textbf{\bibinfo{volume}{23}},
  \bibinfo{pages}{5477} (\bibinfo{year}{2006}), \eprint{gr-qc/0604037}.

\bibitem[{\citenamefont{Pisarski and Jaranowski}(2015)}]{Pisarski:2013dj}
\bibinfo{author}{\bibfnamefont{A.}~\bibnamefont{Pisarski}} \bibnamefont{and}
  \bibinfo{author}{\bibfnamefont{P.}~\bibnamefont{Jaranowski}},
  \bibinfo{journal}{Class. Quant. Grav.} \textbf{\bibinfo{volume}{32}},
  \bibinfo{pages}{145014} (\bibinfo{year}{2015}), \eprint{1302.0509}.

\bibitem[{\citenamefont{Wette}(2014)}]{Wette:2014tca}
\bibinfo{author}{\bibfnamefont{K.}~\bibnamefont{Wette}},
  \bibinfo{journal}{Phys. Rev. D} \textbf{\bibinfo{volume}{90}},
  \bibinfo{pages}{122010} (\bibinfo{year}{2014}), \eprint{1410.6882}.

\bibitem[{\citenamefont{Brown et~al.}(2007)\citenamefont{Brown, Crowder,
  Cutler, Mandel, and Vallisneri}}]{Brown:2007se}
\bibinfo{author}{\bibfnamefont{D.~A.} \bibnamefont{Brown}},
  \bibinfo{author}{\bibfnamefont{J.}~\bibnamefont{Crowder}},
  \bibinfo{author}{\bibfnamefont{C.}~\bibnamefont{Cutler}},
  \bibinfo{author}{\bibfnamefont{I.}~\bibnamefont{Mandel}}, \bibnamefont{and}
  \bibinfo{author}{\bibfnamefont{M.}~\bibnamefont{Vallisneri}},
  \bibinfo{journal}{Class. Quant. Grav.} \textbf{\bibinfo{volume}{24}},
  \bibinfo{pages}{S595} (\bibinfo{year}{2007}), \eprint{0704.2447}.

\bibitem[{\citenamefont{Blaut et~al.}(2009)\citenamefont{Blaut, Krolak, and
  Babak}}]{Blaut:2009zz}
\bibinfo{author}{\bibfnamefont{A.}~\bibnamefont{Blaut}},
  \bibinfo{author}{\bibfnamefont{A.}~\bibnamefont{Krolak}}, \bibnamefont{and}
  \bibinfo{author}{\bibfnamefont{S.}~\bibnamefont{Babak}},
  \bibinfo{journal}{Class. Quant. Grav.} \textbf{\bibinfo{volume}{26}},
  \bibinfo{pages}{204023} (\bibinfo{year}{2009}).

\bibitem[{\citenamefont{Babak}(2008)}]{Babak:2008rb}
\bibinfo{author}{\bibfnamefont{S.}~\bibnamefont{Babak}},
  \bibinfo{journal}{Class. Quant. Grav.} \textbf{\bibinfo{volume}{25}},
  \bibinfo{pages}{195011} (\bibinfo{year}{2008}), \eprint{0801.4070}.

\bibitem[{\citenamefont{Harry et~al.}(2009)\citenamefont{Harry, Allen, and
  Sathyaprakash}}]{Harry:2009ea}
\bibinfo{author}{\bibfnamefont{I.~W.} \bibnamefont{Harry}},
  \bibinfo{author}{\bibfnamefont{B.}~\bibnamefont{Allen}}, \bibnamefont{and}
  \bibinfo{author}{\bibfnamefont{B.~S.} \bibnamefont{Sathyaprakash}},
  \bibinfo{journal}{Phys. Rev. D} \textbf{\bibinfo{volume}{80}},
  \bibinfo{pages}{104014} (\bibinfo{year}{2009}), \eprint{0908.2090}.

\bibitem[{\citenamefont{Fehrmann and Pletsch}(2014)}]{Fehrmann:2014cpa}
\bibinfo{author}{\bibfnamefont{H.}~\bibnamefont{Fehrmann}} \bibnamefont{and}
  \bibinfo{author}{\bibfnamefont{H.~J.} \bibnamefont{Pletsch}},
  \bibinfo{journal}{Phys. Rev. D} \textbf{\bibinfo{volume}{90}},
  \bibinfo{pages}{124049} (\bibinfo{year}{2014}), \eprint{1411.3899}.

\bibitem[{\citenamefont{Allen}(2022)}]{Allen:2022lqr}
\bibinfo{author}{\bibfnamefont{B.}~\bibnamefont{Allen}} (\bibinfo{year}{2022}),
  \eprint{2203.02759}.

\bibitem[{\citenamefont{Manca and Vallisneri}(2010)}]{Manca:2009xw}
\bibinfo{author}{\bibfnamefont{G.~M.} \bibnamefont{Manca}} \bibnamefont{and}
  \bibinfo{author}{\bibfnamefont{M.}~\bibnamefont{Vallisneri}},
  \bibinfo{journal}{Phys. Rev. D} \textbf{\bibinfo{volume}{81}},
  \bibinfo{pages}{024004} (\bibinfo{year}{2010}), \eprint{0909.0563}.

\bibitem[{\citenamefont{Jaranowski and Krolak}(2005)}]{Jaranowski:2005hz}
\bibinfo{author}{\bibfnamefont{P.}~\bibnamefont{Jaranowski}} \bibnamefont{and}
  \bibinfo{author}{\bibfnamefont{A.}~\bibnamefont{Krolak}},
  \bibinfo{journal}{Living Rev. Rel.} \textbf{\bibinfo{volume}{8}},
  \bibinfo{pages}{3} (\bibinfo{year}{2005}), \eprint{0711.1115}.

\bibitem[{\citenamefont{Jaranowski et~al.}(1998)\citenamefont{Jaranowski,
  Krolak, and Schutz}}]{Jaranowski:1998qm}
\bibinfo{author}{\bibfnamefont{P.}~\bibnamefont{Jaranowski}},
  \bibinfo{author}{\bibfnamefont{A.}~\bibnamefont{Krolak}}, \bibnamefont{and}
  \bibinfo{author}{\bibfnamefont{B.~F.} \bibnamefont{Schutz}},
  \bibinfo{journal}{Phys. Rev. D} \textbf{\bibinfo{volume}{58}},
  \bibinfo{pages}{063001} (\bibinfo{year}{1998}), \eprint{gr-qc/9804014}.

\bibitem[{\citenamefont{{Armstrong} et~al.}(1999)\citenamefont{{Armstrong},
  {Estabrook}, and {Tinto}}}]{1999ApJ...527..814A}
\bibinfo{author}{\bibfnamefont{J.~W.} \bibnamefont{{Armstrong}}},
  \bibinfo{author}{\bibfnamefont{F.~B.} \bibnamefont{{Estabrook}}},
  \bibnamefont{and} \bibinfo{author}{\bibfnamefont{M.}~\bibnamefont{{Tinto}}},
  \bibinfo{journal}{\apj} \textbf{\bibinfo{volume}{527}}, \bibinfo{pages}{814}
  (\bibinfo{year}{1999}).

\bibitem[{\citenamefont{Krolak et~al.}(2004)\citenamefont{Krolak, Tinto, and
  Vallisneri}}]{Krolak:2004xp}
\bibinfo{author}{\bibfnamefont{A.}~\bibnamefont{Krolak}},
  \bibinfo{author}{\bibfnamefont{M.}~\bibnamefont{Tinto}}, \bibnamefont{and}
  \bibinfo{author}{\bibfnamefont{M.}~\bibnamefont{Vallisneri}},
  \bibinfo{journal}{Phys. Rev. D} \textbf{\bibinfo{volume}{70}},
  \bibinfo{pages}{022003} (\bibinfo{year}{2004}), \bibinfo{note}{[Erratum:
  Phys.Rev.D 76, 069901 (2007)]}, \eprint{gr-qc/0401108}.

\bibitem[{\citenamefont{Prince et~al.}(2002)\citenamefont{Prince, Tinto,
  Larson, and Armstrong}}]{Prince:2002hp}
\bibinfo{author}{\bibfnamefont{T.~A.} \bibnamefont{Prince}},
  \bibinfo{author}{\bibfnamefont{M.}~\bibnamefont{Tinto}},
  \bibinfo{author}{\bibfnamefont{S.~L.} \bibnamefont{Larson}},
  \bibnamefont{and} \bibinfo{author}{\bibfnamefont{J.~W.}
  \bibnamefont{Armstrong}}, \bibinfo{journal}{Phys. Rev. D}
  \textbf{\bibinfo{volume}{66}}, \bibinfo{pages}{122002}
  (\bibinfo{year}{2002}), \eprint{gr-qc/0209039}.

\bibitem[{\citenamefont{Estabrook et~al.}(2000)\citenamefont{Estabrook, Tinto,
  and Armstrong}}]{Estabrook:2000ef}
\bibinfo{author}{\bibfnamefont{F.~B.} \bibnamefont{Estabrook}},
  \bibinfo{author}{\bibfnamefont{M.}~\bibnamefont{Tinto}}, \bibnamefont{and}
  \bibinfo{author}{\bibfnamefont{J.~W.} \bibnamefont{Armstrong}},
  \bibinfo{journal}{Phys. Rev. D} \textbf{\bibinfo{volume}{62}},
  \bibinfo{pages}{042002} (\bibinfo{year}{2000}).

\bibitem[{\citenamefont{Finn}(1992)}]{Finn:1992wt}
\bibinfo{author}{\bibfnamefont{L.~S.} \bibnamefont{Finn}},
  \bibinfo{journal}{Phys. Rev. D} \textbf{\bibinfo{volume}{46}},
  \bibinfo{pages}{5236} (\bibinfo{year}{1992}), \eprint{gr-qc/9209010}.

\bibitem[{\citenamefont{Cutler and Schutz}(2005)}]{Cutler:2005hc}
\bibinfo{author}{\bibfnamefont{C.}~\bibnamefont{Cutler}} \bibnamefont{and}
  \bibinfo{author}{\bibfnamefont{B.~F.} \bibnamefont{Schutz}},
  \bibinfo{journal}{Phys. Rev. D} \textbf{\bibinfo{volume}{72}},
  \bibinfo{pages}{063006} (\bibinfo{year}{2005}), \eprint{gr-qc/0504011}.

\bibitem[{\citenamefont{Cutler and Flanagan}(1994)}]{Cutler:1994ys}
\bibinfo{author}{\bibfnamefont{C.}~\bibnamefont{Cutler}} \bibnamefont{and}
  \bibinfo{author}{\bibfnamefont{E.~E.} \bibnamefont{Flanagan}},
  \bibinfo{journal}{Phys. Rev. D} \textbf{\bibinfo{volume}{49}},
  \bibinfo{pages}{2658} (\bibinfo{year}{1994}), \eprint{gr-qc/9402014}.

\bibitem[{\citenamefont{Abbott et~al.}(2004)}]{LIGOScientific:2003ijf}
\bibinfo{author}{\bibfnamefont{B.}~\bibnamefont{Abbott}} \bibnamefont{et~al.}
  (\bibinfo{collaboration}{LIGO Scientific}), \bibinfo{journal}{Phys. Rev. D}
  \textbf{\bibinfo{volume}{69}}, \bibinfo{pages}{082004}
  (\bibinfo{year}{2004}), \eprint{gr-qc/0308050}.

\bibitem[{\citenamefont{Abbott et~al.}(2007)}]{LIGOScientific:2006jsu}
\bibinfo{author}{\bibfnamefont{B.}~\bibnamefont{Abbott}} \bibnamefont{et~al.}
  (\bibinfo{collaboration}{LIGO Scientific}), \bibinfo{journal}{Phys. Rev. D}
  \textbf{\bibinfo{volume}{76}}, \bibinfo{pages}{082001}
  (\bibinfo{year}{2007}), \eprint{gr-qc/0605028}.

\bibitem[{\citenamefont{Prix and Whelan}(2007)}]{Prix:2007zh}
\bibinfo{author}{\bibfnamefont{R.}~\bibnamefont{Prix}} \bibnamefont{and}
  \bibinfo{author}{\bibfnamefont{J.~T.} \bibnamefont{Whelan}},
  \bibinfo{journal}{Class. Quant. Grav.} \textbf{\bibinfo{volume}{24}},
  \bibinfo{pages}{S565} (\bibinfo{year}{2007}), \eprint{0707.0128}.

\bibitem[{\citenamefont{Kester}(2003)}]{kester2003mixed}
\bibinfo{author}{\bibfnamefont{W.}~\bibnamefont{Kester}},
  \emph{\bibinfo{title}{Mixed-signal and DSP design techniques}}
  (\bibinfo{publisher}{Newnes}, \bibinfo{address}{Oxford and Boston},
  \bibinfo{year}{2003}).

\bibitem[{\citenamefont{Bracewell}(1986)}]{Bracewell1986The}
\bibinfo{author}{\bibfnamefont{R.~N.} \bibnamefont{Bracewell}},
  \emph{\bibinfo{title}{The Fourier Transofrm and Its Applications, 3rd ed.}}
  (\bibinfo{publisher}{McGraw-Hill}, \bibinfo{address}{New York},
  \bibinfo{year}{1986}).

\bibitem[{\citenamefont{Harry et~al.}(2016)\citenamefont{Harry, Privitera,
  Boh\'e, and Buonanno}}]{Harry:2016ijz}
\bibinfo{author}{\bibfnamefont{I.}~\bibnamefont{Harry}},
  \bibinfo{author}{\bibfnamefont{S.}~\bibnamefont{Privitera}},
  \bibinfo{author}{\bibfnamefont{A.}~\bibnamefont{Boh\'e}}, \bibnamefont{and}
  \bibinfo{author}{\bibfnamefont{A.}~\bibnamefont{Buonanno}},
  \bibinfo{journal}{Phys. Rev. D} \textbf{\bibinfo{volume}{94}},
  \bibinfo{pages}{024012} (\bibinfo{year}{2016}), \eprint{1603.02444}.

\bibitem[{\citenamefont{Allen}(2021)}]{Allen:2021yuy}
\bibinfo{author}{\bibfnamefont{B.}~\bibnamefont{Allen}},
  \bibinfo{journal}{Phys. Rev. D} \textbf{\bibinfo{volume}{104}},
  \bibinfo{pages}{042005} (\bibinfo{year}{2021}), \eprint{2102.11254}.

\bibitem[{\citenamefont{Van Den~Broeck et~al.}(2009)\citenamefont{Van
  Den~Broeck, Brown, Cokelaer, Harry, Jones, Sathyaprakash, Tagoshi, and
  Takahashi}}]{VanDenBroeck:2009gd}
\bibinfo{author}{\bibfnamefont{C.}~\bibnamefont{Van Den~Broeck}},
  \bibinfo{author}{\bibfnamefont{D.~A.} \bibnamefont{Brown}},
  \bibinfo{author}{\bibfnamefont{T.}~\bibnamefont{Cokelaer}},
  \bibinfo{author}{\bibfnamefont{I.}~\bibnamefont{Harry}},
  \bibinfo{author}{\bibfnamefont{G.}~\bibnamefont{Jones}},
  \bibinfo{author}{\bibfnamefont{B.~S.} \bibnamefont{Sathyaprakash}},
  \bibinfo{author}{\bibfnamefont{H.}~\bibnamefont{Tagoshi}}, \bibnamefont{and}
  \bibinfo{author}{\bibfnamefont{H.}~\bibnamefont{Takahashi}},
  \bibinfo{journal}{Phys. Rev. D} \textbf{\bibinfo{volume}{80}},
  \bibinfo{pages}{024009} (\bibinfo{year}{2009}), \eprint{0904.1715}.

\bibitem[{\citenamefont{Indik et~al.}(2017)\citenamefont{Indik, Haris,
  Dal~Canton, Fehrmann, Krishnan, Lundgren, Nielsen, and Pai}}]{RN289}
\bibinfo{author}{\bibfnamefont{N.}~\bibnamefont{Indik}},
  \bibinfo{author}{\bibfnamefont{K.}~\bibnamefont{Haris}},
  \bibinfo{author}{\bibfnamefont{T.}~\bibnamefont{Dal~Canton}},
  \bibinfo{author}{\bibfnamefont{H.}~\bibnamefont{Fehrmann}},
  \bibinfo{author}{\bibfnamefont{B.}~\bibnamefont{Krishnan}},
  \bibinfo{author}{\bibfnamefont{A.}~\bibnamefont{Lundgren}},
  \bibinfo{author}{\bibfnamefont{A.~B.} \bibnamefont{Nielsen}},
  \bibnamefont{and} \bibinfo{author}{\bibfnamefont{A.}~\bibnamefont{Pai}},
  \bibinfo{journal}{Phys. Rev. D} \textbf{\bibinfo{volume}{95}},
  \bibinfo{pages}{064056} (\bibinfo{year}{2017}), \eprint{1612.05173}.

\bibitem[{\citenamefont{Bouffanais and Porter}(2016)}]{Bouffanais:2015sya}
\bibinfo{author}{\bibfnamefont{Y.}~\bibnamefont{Bouffanais}} \bibnamefont{and}
  \bibinfo{author}{\bibfnamefont{E.~K.} \bibnamefont{Porter}},
  \bibinfo{journal}{Phys. Rev. D} \textbf{\bibinfo{volume}{93}},
  \bibinfo{pages}{064020} (\bibinfo{year}{2016}), \eprint{1509.08867}.

\bibitem[{\citenamefont{Wang and Mohanty}(2010)}]{Wang:2010jma}
\bibinfo{author}{\bibfnamefont{Y.}~\bibnamefont{Wang}} \bibnamefont{and}
  \bibinfo{author}{\bibfnamefont{S.~D.} \bibnamefont{Mohanty}},
  \bibinfo{journal}{Phys. Rev. D} \textbf{\bibinfo{volume}{81}},
  \bibinfo{pages}{063002} (\bibinfo{year}{2010}), \eprint{1001.0923}.

\bibitem[{\citenamefont{Taylor et~al.}(2012)\citenamefont{Taylor, Gair, and
  Lentati}}]{Taylor:2012vx}
\bibinfo{author}{\bibfnamefont{S.~R.} \bibnamefont{Taylor}},
  \bibinfo{author}{\bibfnamefont{J.~R.} \bibnamefont{Gair}}, \bibnamefont{and}
  \bibinfo{author}{\bibfnamefont{L.}~\bibnamefont{Lentati}}
  (\bibinfo{year}{2012}), \eprint{1210.3489}.

\bibitem[{\citenamefont{Wang et~al.}(2014)\citenamefont{Wang, Mohanty, and
  Jenet}}]{Wang:2014ava}
\bibinfo{author}{\bibfnamefont{Y.}~\bibnamefont{Wang}},
  \bibinfo{author}{\bibfnamefont{S.~D.} \bibnamefont{Mohanty}},
  \bibnamefont{and} \bibinfo{author}{\bibfnamefont{F.~A.} \bibnamefont{Jenet}},
  \bibinfo{journal}{Astrophys. J.} \textbf{\bibinfo{volume}{795}},
  \bibinfo{pages}{96} (\bibinfo{year}{2014}), \eprint{1406.5496}.

\bibitem[{\citenamefont{Prasad and Souradeep}(2012)}]{Prasad:2011rd}
\bibinfo{author}{\bibfnamefont{J.}~\bibnamefont{Prasad}} \bibnamefont{and}
  \bibinfo{author}{\bibfnamefont{T.}~\bibnamefont{Souradeep}},
  \bibinfo{journal}{Phys. Rev. D} \textbf{\bibinfo{volume}{85}},
  \bibinfo{pages}{123008} (\bibinfo{year}{2012}), \bibinfo{note}{[Erratum:
  Phys.Rev.D 90, 109903 (2014)]}, \eprint{1108.5600}.

\bibitem[{\citenamefont{Prix and Itoh}(2005)}]{Prix:2005gx}
\bibinfo{author}{\bibfnamefont{R.}~\bibnamefont{Prix}} \bibnamefont{and}
  \bibinfo{author}{\bibfnamefont{Y.}~\bibnamefont{Itoh}},
  \bibinfo{journal}{Class. Quant. Grav.} \textbf{\bibinfo{volume}{22}},
  \bibinfo{pages}{S1003} (\bibinfo{year}{2005}), \eprint{gr-qc/0504006}.

\bibitem[{\citenamefont{Littenberg et~al.}(2020)\citenamefont{Littenberg,
  Cornish, Lackeos, and Robson}}]{Littenberg:2020bxy}
\bibinfo{author}{\bibfnamefont{T.}~\bibnamefont{Littenberg}},
  \bibinfo{author}{\bibfnamefont{N.}~\bibnamefont{Cornish}},
  \bibinfo{author}{\bibfnamefont{K.}~\bibnamefont{Lackeos}}, \bibnamefont{and}
  \bibinfo{author}{\bibfnamefont{T.}~\bibnamefont{Robson}},
  \bibinfo{journal}{Phys. Rev. D} \textbf{\bibinfo{volume}{101}},
  \bibinfo{pages}{123021} (\bibinfo{year}{2020}), \eprint{2004.08464}.

\bibitem[{\citenamefont{Vallisneri}(2005)}]{Vallisneri:2004bn}
\bibinfo{author}{\bibfnamefont{M.}~\bibnamefont{Vallisneri}},
  \bibinfo{journal}{Phys. Rev. D} \textbf{\bibinfo{volume}{71}},
  \bibinfo{pages}{022001} (\bibinfo{year}{2005}), \eprint{gr-qc/0407102}.

\end{thebibliography}

\end{document}